\documentclass{aa}

\usepackage{amsmath}
\usepackage{tablefootnote}
\usepackage{graphicx}
\usepackage{graphics}
\usepackage{epsfig}
\usepackage{flushend}
\usepackage{txfonts}
\usepackage{blindtext}
\usepackage[colorlinks=true,linkcolor=blue,citecolor=blue,urlcolor=blue]{hyperref}%

\begin{document}

\title{A quasar hiding behind two dusty absorbers} 
\subtitle{Quantifying the selection bias of metal-rich, damped Ly$\alpha$ absorption systems \thanks{Based on photometric data from the KiDS and VIKING surveys and follow-up observations made with the Gran Telescopio Canarias (GTC), installed in the Spanish Observatorio del Roque de los Muchachos of the Instituto de Astrof\'isica de Canarias, on the island of La Palma.}}

\titlerunning{The dusty absorbers toward KV-RQ\,1500-0013: On the dust bias in DLA samples}

\author{
K.~E.~Heintz\inst{1,2},
J.~P.~U.~Fynbo\inst{2},
C.~Ledoux\inst{3},
P.~Jakobsson\inst{1},
P.~M\o ller\inst{4},
L.~Christensen\inst{2},
S.~Geier\inst{5,6}, \\
J.-K.~Krogager\inst{7}
\and
P.~Noterdaeme\inst{7}
}
\institute{
	Centre for Astrophysics and Cosmology, Science Institute, University of Iceland, Dunhagi 5, 107 Reykjav\'ik, Iceland
	\and
	Dark Cosmology Centre, Niels Bohr Institute, University of Copenhagen, Juliane
	Maries Vej 30, 2100 Copenhagen \O, Denmark
	\email{keh14@hi.is}
	\and
	European Southern Observatory, Alonso de C\'ordova 3107, Vitacura, Casilla 19001, Santiago 19, Chile
	\and
	European Southern Observatory, Karl-Schwarzschildstrasse 2, D-85748 Garching bei M\"unchen, Germany
	\and
	Gran Telescopio Canarias (GRANTECAN), Cuesta de San Jos\'e s/n, E-38712 , Bre\~na Baja, La Palma, Spain 
	\and
	Instituto de Astrof\'isica de Canarias, V\'ia L\'actea s/n, E38200, La Laguna, Tenerife, Spain
	\and
	Institut d'Astrophysique de Paris, CNRS-UPMC, UMR7095, 98bis bd Arago, 75014 Paris, France
}

\authorrunning{Heintz et al.}

\date{Received 2017; accepted, 2017}

\abstract{
The cosmic chemical enrichment as measured from damped Ly$\alpha$ absorbers (DLAs) will be underestimated if dusty and metal-rich absorbers have evaded identification. Here we report the discovery and present the spectroscopic observations of a quasar, KV-RQ\,1500-0031, at $z=2.520$ reddened by a likely dusty DLA at $z=2.428$ and a strong Mg\,\textsc{ii} absorber at $z=1.603$. This quasar was identified as part of the KiDS-VIKING Red Quasar (KV-RQ) survey, specifically aimed at targeting dusty absorbers which may cause the background quasars to escape the optical selection of e.g. the Sloan Digital Sky Survey (SDSS) quasar sample. For the DLA we find an H\,\textsc{i} column density of $\log N$(H\,\textsc{i}) = $21.2\pm 0.1$ and a metallicity of [X/H] = $-0.90\pm 0.20$ derived from an empirical relation based on the equivalent width of Si\,\textsc{ii}\,$\lambda$\,1526. We observe a total visual extinction of $A_V=0.16$ mag induced by both absorbers. To put this case into context we compile a sample of 17 additional dusty ($A_V > 0.1$ mag) DLAs toward quasars (QSO-DLAs) from the literature for which we characterize the overall properties, specifically in terms of H\,\textsc{i} column density, metallicity and dust properties. From this sample we also estimate a correction factor to the overall DLA metallicity budget as a function of the fractional contribution of dusty QSO-DLAs to the bulk of the known QSO-DLA population. We demonstrate that the dusty QSO-DLAs have high metal column densities ($\log N$(H\,\textsc{i}) + [X/H]) and are more similar to gamma-ray burst (GRB)-selected DLAs (GRB-DLAs) than regular QSO-DLAs. We evaluate the effect of dust reddening in DLAs as well as illustrate how the induced color excess of the underlying quasars can be significant (up to $\sim 1$ mag in various optical bands), even for low to moderate extinction values ($A_V \lesssim 0.6$ mag). Finally we discuss the direct and indirect implications of a significant dust bias in both QSO- and GRB-DLA samples.
}

\maketitle

\section{Introduction}

Cosmic lighthouses such as quasars (QSOs) and gamma-ray bursts (GRBs) are powerful tools to study the chemical enrichment of galaxies through cosmic time \citep[e.g.][]{Pettini94,Pettini97b,Pettini97a,Pettini99,Prochaska00,Prochaska03,Prochaska07,Fynbo06,Rafelski12,Cucchiara15}. Damped Ly$\alpha$ absorbers (DLAs) can be identified toward these luminous background sources and are defined as having H\,\textsc{i} column densities of $\log N$(H\,\textsc{i}) > 20.3. The total gas content of DLAs at high redshift is comparable to the visible stellar mass in spiral galaxies at the present epoch and DLAs are thus believed to be the gas reservoirs from which galaxies at high redshift formed \citep{Wolfe86,Wolfe05}. 

On average, DLAs observed in GRB afterglows (GRB-DLAs) have higher H\,\textsc{i} column densities \citep{Vreeswijk04,Jakobsson06,Fynbo09} and metallicities \citep{Savaglio03,Savaglio06,Fynbo06,Prochaska07,Cucchiara15} than DLAs toward quasars (QSO-DLAs). This is expected since QSO-DLAs are cross-section-selected intervening absorbers where the line of sight intersects the galaxy at a random location, whereas GRBs are found to explode in the centremost part or the ultraviolet (UV) brightest regions of the galaxies hosting them \citep{Bloom02,Fruchter06,Lyman17}. As a result, QSO-DLAs will also not sample the luminosity function of star-forming galaxies in the same way \citep{Fynbo08}. However, GRB-DLAs with high H\,\textsc{i} column densities and high metallicities are observed whereas QSO-DLAs appear to be absent in this upper corner of the $Z$\footnote{Where Z is the abundance relative to solar defined as $Z \equiv$ [X/H] = $\log N(\mathrm{X})/N(\mathrm{H}) - \log N(\mathrm{X}/\mathrm{H})_{\odot}$.}$-N$(H\,\textsc{i}) plane. \cite{Boisse98} argued that this apparent threshold could imply a significant dust bias in current, optically selected QSO-DLA samples.
This bias, if confirmed, will cause an underestimation of the metallicity, the contribution of neutral gas to the cosmological mass density and the estimated global molecular fraction. 
The extent to which dusty and therefore also metal-rich QSO-DLAs might be under-represented in optically selected quasar samples due to dust obscuration has been studied extensively in the literature \citep[see e.g.][]{Pei91,Fall93,Vladilo05,Smette05,Trenti06,Pontzen09} but where some claim it to be negligible, other find it to be significant.

In addition to dust obscuration (simply dimming the underlying quasar) the effect of reddening of otherwise bright quasars will also play an important role in the detection probability of these systems. As an example, \cite{Fynbo11} discovered a metal-rich DLA on the verge of dropping out of the optical quasar selection of the Sloan Digital Sky Survey \citep[SDSS;][]{Richards02,Richards04} framework. This initiated the High $A_V$ Quasar \citep[HAQ;][]{Fynbo13a,Krogager15,Krogager16b} survey, tailored for the detection of dusty, intervening QSO-DLAs. Two dusty DLAs have so far been identified in this survey, both toward quasars misclassified as stars in the SDSS \citep{Krogager16a,Fynbo17}. The large majority of the sample, however, consists of intrinsically dust-reddened quasars and quasars likely reddened by absorbers at $z<2$. Some evidence for a dust-bias in optical samples has indeed also been found from radio-selected QSO-DLAs \citep[identifying quasars at radio wavelengths should not be affected by foreground dust, e.g;][]{Ellison01,Ellison08,Akerman05,Jorgenson06,Ellison09}, though so-far only from small-number studies, which are therefore not fully conclusive. 

After completion of the HAQ survey we redefined our search for the elusive, dusty QSO-DLAs through a new campaign designated the KiDS-VIKING Red Quasar (KV-RQ) survey (Heintz et al., in preparation). This survey was built from the Kilo Degree Survey \citep[KiDS;][]{deJong13,deJong15} and the VISTA Kilo-degree Infrared Galaxy \citep[VIKING;][]{Edge13} photometric data, which are roughly two magnitudes deeper than the SDSS and the UKIRT Infrared Deep Sky Survey (UKIDSS) detection limits that the HAQ survey relied on. Furthermore, after the large number of dust-reddened quasars that have now been classified, we were able to increase the efficiency of identifying quasars at $z>2$ based on optical to near/mid-infrared photometry only and at the same time remove most of the previously observed contaminants (e.g., dwarf stars and compact galaxies). 

In this paper we present the third detection of a quasar, for short named KV-RQ\,1500-0013, reddened by a dusty and metal-rich DLA and an intervening Mg\,\textsc{ii} absorber identified as part of the KV-RQ survey. Similar to the other two detections in the HAQ survey this source is classified as a star in the SDSS survey, as well as in the Baryon Oscillation Spectroscopic Survey (BOSS) program \citep{Ross12,Myers15}, which nevertheless relies on more complex selection algorithms. It was subsequently selected for spectroscopic follow-up observations by us as it fulfils the specific optical to near/mid-infrared photometric selection criteria defined for the KV-RQ survey (Heintz et al., in preparation). With this third detection the substantial number of dusty DLAs that have been reported in the literature, we proceed to evaluate and characterize these elusive DLAs as a population. Specifically, we test how these dusty QSO-DLAs compare to regular QSO-DLAs and GRB-DLAs in terms of H\,\textsc{i} column density, metallicity (and thus also the metal column density), and dust properties. Moreover, we want to quantify empirically the effect of reddening due to dust on the background quasar. This will allow us to better assess the effect of a dust bias in existing optically selected quasar samples. 

The paper is structured as follows. In Section~\ref{sec:obs} we present our new spectroscopic observations of the quasar KV-RQ\,1500-0013. In Section~\ref{sec:abs} we report the results of analyzing the quasar and its intervening DLA and Mg\,\textsc{ii} absorber, with special focus on the DLA. Section~\ref{sec:dlacomp} is dedicated to the evaluation and the characteristics of the population of dusty QSO-DLAs as a whole, combining this new observation with previous detections reported in the literature. Here we will also compare this subset of elusive QSO-DLAs to the bulk of known QSO-DLAs and the population of GRB-DLAs. In Section~\ref{sec:bias}, we discuss the dust bias induced specifically in optically selected quasar samples, by the significant reddening effect these dusty absorbers have on the optical colors of the underlying quasars. Finally in Section~\ref{sec:conc} we summarize and conclude our work with emphasis on the implications of dust bias in DLA samples.

Throughout the paper we assume concordance cosmology with $\Omega_m=0.315$, $\Omega_{\Lambda}=0.685$, and $H_0=67.3$ km s$^{-1}$ Mpc$^{-1}$ \citep{Planck14}, the solar abundances of \cite{Asplund09} and report magnitudes in the AB \citep{Oke83} magnitude system. 

\section{Observations} \label{sec:obs}

We observed KV-RQ\,1500-0013 with the Optical System for Imaging and low-intermediate-Resolution Integrated Spectroscopy (OSIRIS) instrument mounted at the Gran Telescopio Canarias (GTC) as part of a larger sample, specifically aimed at targeting candidate quasars reddened by dusty DLAs. The full sample paper describing the selection strategy and sample properties will be released once the full campaign has been completed. The DLA and Mg\,\textsc{ii} absorber were identified after securing longslit spectroscopy with OSIRIS using the R1000B grism covering $3630 - 7500$\,\AA. To better constrain the spectral energy distribution (SED) and the metal line measurements we obtained follow-up observations with the higher resolution R2500V and R2500R grisms, covering the spectral ranges $4500 - 6000$\,\AA~and $5575 - 7685$\,\AA, respectively. The log of observations at GTC is provided in Table~\ref{tab:obs}.

\begin{table}[!h]
	\centering
	\begin{minipage}{\columnwidth}
		\centering
		\caption{Log of observations with the GTC \label{tab:obs}}
		\begin{tabular}{lcccc}
			\noalign{\smallskip} \hline \hline \noalign{\smallskip}
			Date & Grism & $\mathcal{R}$ &  Exp. time (s) & Airmass \\
			\noalign{\smallskip}\hline \noalign{\smallskip}
				04-06-2017 & R1000B & 500 & $3\times 450$ & 1.37 \\
				22-06-2017 & R2500V & 1500 & $3\times 1000$ & 1.39 \\
				17-07-2017 & R2500R & 1850 & $3\times 1000$ & 1.31 \\
			\noalign{\smallskip}\hline \noalign{\smallskip}
		\end{tabular}
		\centering
		\tablefoot{The specific coverage and instrumental resolution of the used grisms can be found at the OSIRIS webpage\tablefootnote{\url{http://www.gtc.iac.es/instruments/osiris/\#Longslit_Spectroscopy}}, the naming here following the denotation listed there.
		}
	\end{minipage}
\end{table}

\begin{figure*} [ht!]
	\centering
	\epsfig{file=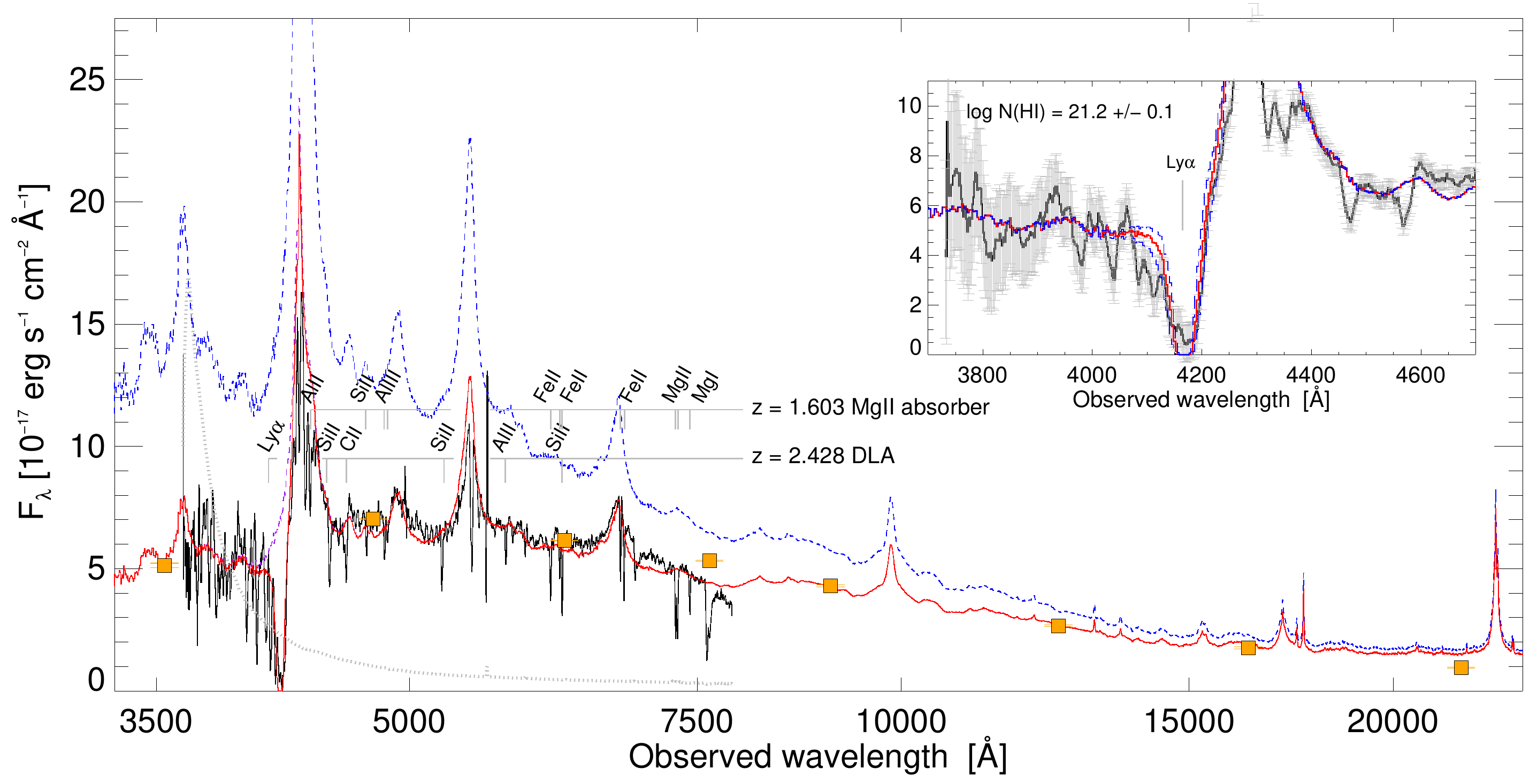,width=0.9\textwidth}
	\caption{GTC spectrum taken with the R1000B grism shown as the solid black line together with the optical/near-infrared photometry in the $u,~g,~r,~i,~Z,~Y,~J,~H,$ and $K_s$ bands from KiDS and VIKING overplotted as orange squares. The error spectrum is shown as the gray dotted line. In blue is shown a composite quasar template and in red the same composite but reddened by $A_V=0.16$ mag in the rest frame of the DLA with the model DLA imposed on the template. The purple line shows the reddened template without the DLA. In the inset we zoom in on the spectral region covering the Ly$\alpha$ absorption line of the DLA at $z_{\mathrm{DLA}}=2.4284$ and show the model with the best-fit H\,\textsc{i} column density and associated error of $\log N$(H\,\textsc{i}) = $21.2\pm 0.1$ as the red solid and blue dashed lines respectively.  \label{fig:spec}}
\end{figure*}

The spectra were reduced using standard procedures for low-resolution spectroscopy in IRAF\footnote{IRAF is distributed by the National Optical Astronomy Observatory, which is operated by the Association of Universities for Research in Astronomy (AURA) under a cooperative agreement with the National Science Foundation.} after the cosmic rays were rejected using the software written by \cite{vanDokkum01}. The spectra were flux calibrated using observations of spectrophotometric standard stars obtained on the same night as the science spectra. Both the spectra and the photometry were corrected for Galactic extinction using the dust maps from \cite{Schlafly11}. All wavelengths reported are in vacuum and are corrected to the heliocentric rest frame. To improve the absolute flux calibration and correct for any slit-loss effects we scaled the observed spectra to the $r$-band photometry from KiDS.

\section{Results} \label{sec:abs}

We present the GTC spectrum taken with the R1000B grism in Fig.~\ref{fig:spec} together with the optical/near-infrared photometry from KiDS  and VIKING. Overplotted in blue is a composite quasar template \citep{Selsing16}, where in red we show the same composite but reddened by a visual extinction of $A_V=0.16$ mag (see below). The advantage of this composite is that it was created from observations of luminous blue quasars observed with the Very Large Telescope (VLT)/X-shooter which provides a broad spectral coverage without any significant host contamination and ensures that the template is constructed from quasars observed in a consistent manner. It is clear that the overall shape of the spectrum is consistent with having a substantial amount of reddening. We identify two intervening absorbers in the spectrum of the quasar at $z=2.520$, a DLA at a systemic redshift of $z=2.4284\pm 0.0004$, and a Mg\,\textsc{ii} absorber at $z=1.6027 \pm 0.0005$. We will treat the analysis of the two absorbers separately, Sect.~\ref{ssec:dla} being dedicated to the DLA and Sect.~\ref{ssec:mgii} to the Mg\,\textsc{ii} absorber. In Sect.~\ref{ssec:ext} we discuss the likely contribution of the two absorbers to the total observed extinction.

\subsection{The DLA at $z=2.428$} \label{ssec:dla}

\subsubsection{Neutral hydrogen column density}

The Ly$\alpha$ absorption line was detected in the first observation using the R1000B grism, covering the full spectral region of this feature. In the inset in the upper right corner of Fig.~\ref{fig:spec} we show a zoom-in of this region. The H\,\textsc{i} column density was derived by fitting the reddened composite quasar template with a DLA included in the model \citep[using the approximation of][]{TepperGracia06} to the observed spectrum. From the best-fit model we find an H\,\textsc{i} column density of $\log N$(H\,\textsc{i}) = $21.2\pm 0.1$ at the redshift $z_{\mathrm{DLA}}=2.428$. Due to the small discrepancy between the model and the continuum spectrum on the blue side of the Ly$\alpha$ trough, we normalized the region around the absorption line and fit the column density independently. We found the H\,\textsc{i} column density to be consistent with that derived from the reddened quasar composite with the imposed DLA model.

Even though located close in redshift to the quasar at $z_{\mathrm{QSO}}=2.520$, we note that this DLA does not belong to the subclass of proximate DLAs (PDLAs), typically defined to be less than 3000 or 5000 km s$^{-1}$ from the quasar emission redshift \citep{Ellison02,Russell06,Prochaska08b}, where we measure $\Delta(z_{\mathrm{QSO}}-z_{\mathrm{DLA}}) \approx 8000$ km s$^{-1}$. Due to the close proximity of PDLAs to the background quasar, they have in general different characteristics to interving DLAs and might be associated with the quasar host galaxy. There are also indications of enhanced ionization in PDLAs causing the overall properties of these to contrast with those of typical intervening DLAs \citep{Moller98a,Prochaska08b,Ellison10}.

\subsubsection{Metallicity}

\begin{table}[!h]
	\centering
	\begin{minipage}{\columnwidth}
		\centering
		\caption{Detected absorption lines and rest-frame equivalent widths of the metal lines associated with the DLA and the Mg\,\textsc{ii} absorber
		\label{tab:ewabs}}
		\begin{tabular}{lccc}
			\noalign{\smallskip} \hline \hline \noalign{\smallskip}
			Transition & $\lambda_{\mathrm{obs}}$ & $z$ &  $W_{\mathrm{rest}}$ \\
			& (\AA) & & (\AA)    \\
			\noalign{\smallskip}\hline \noalign{\smallskip}
			DLA  &&& \\
			\noalign{\smallskip}\hline \noalign{\smallskip}
			O\,\textsc{i}~$\lambda$~1302 & 4464.10 & 2.4282 & $0.65\pm 0.08$  \\
			Si\,\textsc{ii}~$\lambda$~1304 & 4471.58 & 2.4282 & $0.57\pm 0.08$    \\
			C\,\textsc{ii}~$\lambda$~1334 & 4574.82 & 2.4280 & $0.81\pm 0.07$  \\
			Si\,\textsc{ii}~$\lambda$~1526 & 5234.08 & 2.4283 & $1.06\pm 0.02$ \\
			C\,\textsc{iv}~$\lambda$~1548* & 5308.18 & 2.4286 & $0.15\pm 0.03$  \\
			Fe\,\textsc{ii}~$\lambda$~1608 & 5515.01 & 2.4288 & $0.50\pm 0.02$  \\
			Al\,\textsc{ii}~$\lambda$~1670 & 5728.74 & 2.4288 & $0.77\pm 0.10$  \\
			Si\,\textsc{ii}~$\lambda$~1808* & 6200.71 & 2.4296 & $1.82\pm 0.02$  \\
			\noalign{\smallskip}\hline \noalign{\smallskip}
			Mg\,\textsc{ii} absorber &&& \\
			\noalign{\smallskip}\hline \noalign{\smallskip}
			Al\,\textsc{ii}~$\lambda$~1670 & 4347.65 & 1.6022 & $1.70\pm 0.18$ \\
			Si\,\textsc{ii}~$\lambda$~1808 & 4706.66 & 1.6032 & $1.26\pm 0.13$  \\
			Al\,\textsc{iii}~$\lambda$~1854 & 4826.14 & 1.6021 & $1.20\pm 0.08$ \\
			Al\,\textsc{iii}~$\lambda$~1862 & 4847.12 & 1.6021 & $0.84\pm 0.07$  \\
			Zn\,\textsc{ii}~$\lambda$~2026 & 5273.35 & 1.6027 & $0.45\pm 0.03$ \\
			Zn\,\textsc{ii}~$\lambda$~2062 & 5368.27 & 1.6026 & $0.19\pm 0.02$ \\
			Fe\,\textsc{ii}~$\lambda$~2344 & 6101.29 & 1.6027 & $1.97\pm 0.04$  \\
			Fe\,\textsc{ii}~$\lambda$~2374 & 6180.07 & 1.6027 & $1.21\pm 0.03$  \\
			Fe\,\textsc{ii}~$\lambda$~2382* & 6201.48 & 1.6026 & $2.30\pm 0.04$  \\
			Mn\,\textsc{ii}~$\lambda$~2576 & 6707.60 & 1.6030 & $0.22\pm 0.02$ \\
			Fe\,\textsc{ii}~$\lambda$~2586 & 6732.68 & 1.6029 & $1.84\pm 0.02$  \\
			Mn\,\textsc{ii}~$\lambda$~2594 & 6754.18 & 1.6033 & $0.21\pm 0.02$ \\
			Fe\,\textsc{ii}~$\lambda$~2600 & 6767.70 & 1.6028 & $2.47\pm 0.02$  \\
			Mn\,\textsc{ii}~$\lambda$~2606 & 6783.95 & 1.6027 & $0.21\pm 0.02$ \\
			Mg\,\textsc{ii}~$\lambda$~2796 & 7279.00 & 1.6038 & $3.14\pm 0.06$  \\
			Mg\,\textsc{ii}~$\lambda$~2803 & 7297.43 & 1.6037 & $3.12\pm 0.08$  \\
			Mg\,\textsc{i}~$\lambda$~2852 & 7426.63 & 1.6039 & $1.45\pm 0.03$  \\
			\noalign{\smallskip}\hline \noalign{\smallskip}
		\end{tabular}
		\centering
		\tablefoot{*Blended.}
	\end{minipage}
\end{table}

The metal lines associated with the DLA are listed in Table~\ref{tab:ewabs} where we also report the measured redshifts and the rest-frame equivalent widths ($W_{\mathrm{rest}}$) for each of the lines. From the low-ionization absorption features we find the DLA to be located at a systemic redshift of $z=2.4284\pm 0.0004$. To determine $W_{\mathrm{rest}}$ we fitted the continuum around each of the lines in regions that were free of absorption features and tellurics. We then summed over the absorption profile contained below the normalized flux level. All the observed features are heavily saturated. The medium resolution of the spectra ($\mathcal{R} \approx 160 - 200$ km s$^{-1}$ in the R2500V- and R2500R-grism observations) is too low to allow for a proper Voigt-profile fit to the absorption lines. For example, from high-resolution spectra of GRB afterglows, \cite{Prochaska06} have shown that strong metal absorption lines usually consists of a number of narrow ($\leq 20$ km s$^{-1}$), unsaturated components. We are therefore not able to derive a reliable column density based on Voigt-profile fitting. Furthermore, Si\,\textsc{ii}\,$\lambda$\,1808 is blended with Fe\,\textsc{ii}\,$\lambda$\,2382 from the Mg\,\textsc{ii} absorber (see Table~\ref{tab:ewabs}), which also rules out the possibility of deriving a reliable column density from a single-ion curve-of-growth technique.

To compute the metallicity of the DLA we have to instead rely on the empirical relation by \cite{Prochaska08a}. They found that the gas-phase metallicity, [X/H], is tightly correlated with the rest-frame equivalent width of Si\,\textsc{ii}\,$\lambda$\,1526. This correlation is argued to be a representation of the mass-metallicity relation in QSO- and GRB-DLAs, since Si\,\textsc{ii}\,$\lambda$\,1526 traces the dynamical motions of the halo gas outside the interstellar medium (ISM), thus probing the total mass contained within the galaxy halo. Based on their relation we find [X/H] = $-0.90 \pm 0.20$ for the DLA, where both the errors from the measured rest frame $W_{\mathrm{1526}}$ and from the scatter in the relation are included. We caution that since this metallicity estimate is based solely on an empirical relation, the actual metallicity might be different and can only be recovered from higher resolution spectra. Furthermore, the true metallicity could be higher due to the depletion of Si onto dust grains.

\subsection{A strong Mg\,II absorber at $z=1.603$} \label{ssec:mgii}

In addition to the DLA at $z=2.428$ we identify an intervening Mg\,\textsc{ii} absorber in the quasar spectrum as well. We detected several metal lines associated with the absorber and list them in Table~\ref{tab:ewabs}, where again the measured redshifts and the rest-frame equivalent widths for each of the lines are reported. From the low-ionization absorption lines we find a systemic redshift of $z=1.6027\pm 0.0005$ of the Mg\,\textsc{ii} absorber. The equivalent widths have been measured in the same way as for the DLA. This Mg\,\textsc{ii} absorber is remarkably strong, where the measured rest-frame equivalent widths of the two components of the Mg\,\textsc{ii} doublet are $W_{\mathrm{2796}}=3.14\pm 0.06$\,{\AA}~and $W_{\mathrm{2803}}=3.12\pm 0.08$\,{\AA}, respectively. 

Compared to the two samples of Mg\,\textsc{ii} absorbers, one for intervening systems toward quasars and one for absorbers toward GRBs, compiled by \cite{Christensen17}, the strength of $W_{\mathrm{2796}}$ is among the 7\% strongest Mg\,\textsc{ii} absorbers for any given path length of $\Delta z = 1$. Comparing this system to the extensive sample of DLAs automatically identified in the SDSS survey \citep{Noterdaeme12b}, we find that only $\approx15\%$ of the intervening Mg\,\textsc{ii} absorbers have $W_{\mathrm{2796}}>3$\,{\AA}, establishing this particular system as a relatively rare Mg\,\textsc{ii} absorber with such strong features imprinted in absorption. 

There is indirect evidence for this Mg\,\textsc{ii} absorber being a DLA as well. However, since the Ly$\alpha$ absorption feature is located at $\approx 3150$\,\AA~at this redshift, it falls just outside the spectral coverage of the R1000B and R2500U grisms available for OSIRIS at GTC so that we are unable to confirm the high column density of neutral hydrogen. \cite{Rao06} found from their survey of Mg\,\textsc{ii} absorbers at $z < 1.65$ that systems that are classified as DLAs are confined to the region where the ratio between the rest-frame Mg\,\textsc{ii} and Fe\,\textsc{ii} equivalent widths is $1 \lesssim W_{2796}/W_{2600} \lesssim 2$. We measure a ratio of $W_{2796}/W_{2600} = 1.27\pm 0.03$ in our case. Furthermore, all of the absorbers from their sample with $W_{2600}>2$\,\AA~have $\log N$(H\,\textsc{i})~$>20$ and for $W_{2852}>1$\,\AA~have $\log N$(H\,\textsc{i})~$>20.6$. We measure $W_{2600} = 2.47\pm 0.02$\,\AA~and $W_{2852} = 1.45\pm 0.03$\,\AA, respectively.

\subsection{Extinction} \label{ssec:ext}

To derive the amount of extinction in the sightline towards the quasar, KV-RQ\,1500-0013, we assume that the composite quasar template of \cite{Selsing16} is a good representation of the underlying, intrinsic shape of the spectrum. While it is not possible to unambiguously know the exact shape, a large majority of the brightest quasars do, however, appear to have similar spectra which also result in the similarities found between individual composite quasar spectra \citep{Francis91,VandenBerk01,Telfer02,Glikman06,Lusso15,Selsing16}.

Specifically, we fit the model of a composite quasar template with an imposed DLA at the derived redshift to the observed spectrum and photometry. In the majority of sighlines toward QSO- and GRB-DLAs the wavelength dependence on dust extinction is well reproduced by that derived toward individual stars in the Milky Way (MW) and the Small and Large Magellanic Clouds \citep[SMC and LMC; e.g.][]{Schady10,Covino13,Japelj15,Ledoux15,Krogager15,Krogager16b}. In these sightlines, the dust model derived by \cite{Pei92} is parametrized as
\begin{equation}
A_{\lambda} = A_V \left(\frac{1}{R_V} + 1\right) \sum_{i=1}^{6} \frac{a_i}{(\lambda/\lambda_i)^{n_i}+(\lambda_i/\lambda)^{n_i}+b_i}~,
\end{equation}
where the parameters $a_i$, $b_i$ and $n_i$ are different for each reddening law, and also the total-to-selective extinction ratio, $R_V = 3.08$ (MW), $R_V = 2.93$ (SMC) and $R_V = 3.16$ (LMC). In addition, we also used the steeper extinction curve ($R_V = 2.41$) derived by \cite{Zafar15} for a sub-sample of intrinsically dust-reddened quasars from the HAQ survey. 
We fit the observed data to all of the four extinction curves, where the dust can be located at the three different redshifts; $z_{\mathrm{MgII}}$, $z_{\mathrm{DLA}}$ and $z_{\mathrm{QSO}}$. We found a best fit of $A_V=0.16$ mag assuming the SMC extinction curve, shown in Fig.~\ref{fig:spec} as the red lines with (solid) and without (dashed) the DLA imposed on the model. We note, however, that using the steeper extinction curve by \cite{Zafar15} leads to consistent fits only with a lower best-fit value of $A_V$ as also shown by \cite{Krogager16b}. Since both of the intervening absorbers are good candidates for containing a significant amount of dust (see below), we argue that the dust is unlikely to be intrinsic to the quasar. So, the extinction curve of \cite{Zafar15} is likely to be not a good assumption in our case and we therefore assume the SMC extinction curve throughout the paper. 

We are unable to disentangle the exact location of the dust component. Only in cases where the 2175\,\AA~extinction bump feature is detected is it possible to determine the redshift of the dust imprinting this feature on the observed spectrum \citep[see, e.g.,][for detections of the 2175\,\AA~extinction bump in the sightlines toward QSO- and GRB-DLAs]{Srianand08,Eliasdottir09,Prochaska09,Noterdaeme09b,Jiang11,Kulkarni11,Wang12,Zafar12,Ma15,Fynbo17}. Even the strength of the 2175\,\AA~feature can be determined \citep[e.g.][]{Jiang10,Ledoux15}. These authors found that heavy dust depletion is required to produce a pronounced 2175\,\AA~extinction bump and that there also appears to be a tight correlation between the strength of the bump and the amount of extinction. 

We do not detect any indication of a bump in our case, thereby excluding the MW and LMC extinction curves, and can therefore not robustly conclude on the location of the dust and the amount of dust located at the different redshifts. The best-fit extinction, however, is consistent with the relatively large metal column density of the DLA system (see e.g. Sect.~\ref{ssec:dladust}), where we expect  $A_V=0.13^{+0.13}_{-0.06}$ mag based on the relation from \cite{Zafar13}, including the scatter. On the other hand, comparing the rest-frame equivalent widths of Fe\,\textsc{ii}\,$\lambda$\,1608 from the DLA and Fe\,\textsc{ii}\,$\lambda$\,2374 from the Mg\,\textsc{ii} absorber (these two transitions have similar oscillator strengths) indicates that the Mg\,\textsc{ii} system is a factor of two stronger than the DLA in terms of metal content and/or velocity dispersion. We note, however, that for a given amount of dust, stronger reddening effects will be produced at higher redshifts due to the shape of the extinction curve. The observed spectral shape of the quasar is thus more sensitive to dust located in the DLA. Based on the metal lines, however, the Mg\,\textsc{ii} absorber might be the best candidate for hosting the largest amount of dust.

\begin{table*}[!ht]
	\centering
	\begin{minipage}{\textwidth}
		\centering
		\caption{Properties of the compiled sample of dusty QSO-DLAs}
		\begin{tabular}{lccccccclc}
			\noalign{\smallskip} \hline \hline \noalign{\smallskip}
			Source & $z_{\mathrm{QSO}}$ &  $z_{\mathrm{DLA}}$ & $A_V$ & $\log N$(H\,\textsc{i}) & [X/H] & X & [Fe/X] & Survey & References  \\
			&       &    & (mag) & (cm$^{-2}$)&&&&&   \\
			\noalign{\smallskip} \hline \noalign{\smallskip}
			J\,0000+0048 & 3.028 & 2.525 & 0.23 & $20.80\pm 0.10$ & $~~0.46\pm 0.45$ & Zn & $-1.89\pm 0.50$ & BOSS\tablefootmark{a,b} & (1) \\
			J\,0016-0012 & 2.092 & 1.970 & 0.11 & $20.75\pm 0.15$ & $-0.86\pm 0.13$ & $\cdots^*$ & $-0.99\pm 0.10$ & BOSS\tablefootmark{a} &(2) \\
			eHAQ\,0111+0641 & 3.214 & 2.027 & 0.22 & $21.50\pm 0.30$ & $-0.60\pm 0.30$ & Zn & $-0.60\pm 0.10$ & eHAQ\tablefootmark{c} & (3,4) \\
			J\,0316+0040 & 2.911 & 2.180 & 0.12 & $21.10\pm 0.20$ & $-1.20\pm 0.22$ & $\cdots^*$ & $-0.63\pm 0.20$ & SEGUE2\tablefootmark{d} & (2) \\
			J\,0812+3208 & 2.711 & 2.626 & 0.11 & $21.35\pm 0.15$ & $-0.88\pm 0.11$ & $\cdots^*$ & $-0.15\pm 0.05$ & BOSS\tablefootmark{a,e} & (2) \\
			J\,0816+1446 & 3.847 & 3.287 & 0.15 & $22.00\pm 0.10$ & $-1.10\pm 0.10$ & Zn & $-0.48\pm 0.02$ & BOSS\tablefootmark{a} & (5) \\
			J\,0840+4942 & 2.090 & 1.851 & 0.13 & $20.75\pm 0.15$ & $-0.43\pm 0.17$ & $\cdots^*$ & $-0.71\pm 0.15$ & SDSS\tablefootmark{$\star$,f} & (2) \\
			Q\,0918+1636 & 3.070 & 2.583 & 0.22 & $20.96\pm 0.05$ & $-0.12\pm 0.05$ & Zn & $-0.91\pm 0.08$ & BOSS\tablefootmark{$\star$,a,b,g} & (6) \\
			J\,0927+5823 & 1.910 & 1.635 & 0.38 & $20.40\pm 0.25$ & $-0.22\pm 0.25$ & $\cdots^*$ & $\cdots$ & BOSS\tablefootmark{e} & (2) \\
			J\,1135-0010 & 2.888 & 2.207 & 0.11 & $22.10\pm 0.05$ & $-1.10\pm 0.20$  & Zn & $-0.72\pm 0.10$ & BOSS\tablefootmark{$\star$,a,g,i,j} & (7) \\
			J\,1157+6155 & 2.513 & 2.460 & 0.59 & $21.80\pm 0.20$ & $-0.60\pm 0.38$ & Zn & $-1.02\pm 0.47$ & SDSS\tablefootmark{$\star$,h} & (8,9) \\
			J\,1211+0833 & 2.483 & 2.117 & 0.53 & $21.00\pm 0.20$ & $-0.07\pm 0.21$ & Zn & $-1.74\pm 0.25$ & BOSS\tablefootmark{$\star$,g,j} & (10) \\
			J\,1310+5424 & 1.929 & 1.801 & 0.27 & $21.45\pm 0.15$ & $-0.51\pm 0.15$ & $\cdots^*$ & $\cdots$ & SDSS\tablefootmark{$\star$,f,h} & (2)  \\
			KV-RQ\,1500-0013 & 2.520 & 2.428 & 0.16 & $21.20\pm 0.10$ & $-0.90\pm 0.20$ & $\cdots^*$ & $\cdots$ & KV-RQ\tablefootmark{c} & This work \\
			J\,1709+3258 & 1.889 & 1.830 & 0.17 & $20.95\pm 0.15$ & $-0.28\pm 0.15$ & $\cdots^*$ & $-0.53\pm 0.15$ & BOSS\tablefootmark{k} & (2) \\
			Q\,2222+0946 & 2.927 & 2.354 & 0.15 & $20.55\pm 0.15$ & $-0.46\pm 0.07$ & Zn & $-0.77\pm 0.05$ & SDSS\tablefootmark{$\star$,f,l} & (2,11) \\
			HAQ\,2225+0527 & 2.320 & 2.130 & 0.28 & $20.69\pm 0.05$ & $-0.09\pm 0.05$ & Zn & $-1.22\pm 0.06$ & HAQ\tablefootmark{c} & (12,13) \\
			J\,2340-0053 & 2.084 & 2.055 & 0.19 & $20.35\pm 0.15$ & $-0.59\pm 0.15$ & $\cdots^*$ & $-0.53\pm 0.12$ & BOSS\tablefootmark{e} & (2,8)  \\
			\noalign{\smallskip} \hline \noalign{\smallskip}
		\end{tabular}
		\tablefoot{
			Selection flags given in the SDSS database: \\
			\tablefoottext{a}{Known QSO at $z>2.15$;} \tablefoottext{b}{Additional object observed to reach 40 QSOs deg$^{-1}$ for BOSS outside the standard selection criteria;} \tablefoottext{c}{Red optical to near/mid-infrared selection, classified as star in SDSS/BOSS;} \tablefoottext{d}{Classified as a star in the SEGUE-2 survey;} \tablefoottext{e}{FIRST radio selected;} \tablefoottext{f}{Bright $ugri$-selected quasar;} \tablefoottext{g}{Identified by the neural network and the kernel density estimation (KDE) plus $\chi^2$ algorithms of BOSS;} \tablefoottext{h}{High-redshift $griz$-selected quasar;} \tablefoottext{i}{Identified by the BOSS likelihood method;} \tablefoottext{j}{Selected to be in the uniform BOSS core sample;} \tablefoottext{k}{Filler target;} \tablefoottext{l}{Classified as a blue horizontal branch star as part of the SEGUE target selection}. \\
			$^{\star}$Quasars selected from the photometric selection criteria and identification algorithms defined specifically for the SDSS/BOSS surveys. \\
			*Metallicities from \citet{Kaplan10} are either based on Si\,\textsc{ii}, Zn\,\textsc{ii} or S\,\textsc{ii}, but the exact ion is not listed in their table 2, and for KV-RQ\,1500-0013 the metallicity is derived from an empirical relation based on the equivalent width of Si\,\textsc{ii}\,$\lambda$\,1526.
		}
		\tablebib{
			(1)~\citet{Noterdaeme17}; (2)~\citet{Kaplan10}; (3)~\citet{Krogager16b}; (4)~\citet{Fynbo17}; (5)~\citet{Guimaraes12}; (6)~\citet{Fynbo11}; (7)~\citet{Noterdaeme12a}; (8)~\citet{Ledoux15}; (9)~\cite{Wang12}; (10)~\citet{Ma15}; (11)~\cite{Fynbo10}; (12)~\citet{Krogager15}; (13)~\citet{Krogager16a}.
		}
		\centering
		\label{tab:log}
	\end{minipage}
\end{table*}

\section{Properties of dusty QSO-DLAs} \label{sec:dlacomp}

With this now third detection of a quasar likely to be reddened by a dusty DLA (though we cannot fully rule out the possibility of a shared dust contribution from, for example, the Mg\,\textsc{ii} absorber) from our surveys, misclassified as a star in SDSS, we will evaluate both the characteristics of these elusive absorbers and the quasar selection methods through which the background quasars are selected. So far only a small population of dusty QSO-DLAs have been analyzed \citep[see e.g.][and the references listed in Table~\ref{tab:log}]{Vladilo06}. Characterizing this dusty subset of the overall population of QSO-DLAs and understanding the selection bias toward this sub-population are of crucial importance in probing the full underlying population of DLAs (and in general also other types of quasar absorbers) and thus also estimating the true cosmic metallicity distribution. 

In addition to the dusty DLA towards KV-RQ\,1500-0013, we compiled a sample of similar absorption systems by extracting several dusty QSO-DLAs from the literature where we required the following criteria: \textit{i}) $\log N$(H\,\textsc{i}) > 20.3 cm$^{-2}$, \textit{ii}) have (at least) optical magnitudes from the SDSS, \textit{iii}) existing metallicity measurements from any of the mildly depleted elements (preferably Zn, otherwise S) and \textit{iv}) visual extinction larger than $A_V\geq 0.1$ mag. Our final sample consists of 18 dusty QSO-DLAs and is presented in Table~\ref{tab:log}. The first criterion is to ensure that the absorber can be classified as a DLA and the second is required so that we can examine the reddening effects on the background QSO (see Sect.~\ref{sec:bias}). The third criterion is defined so that we can explore the distribution of these dusty QSO-DLAs in the metallicity versus H\,\textsc{i} column density plane. Specifically we wish to examine if the apparent upper envelope found for QSO-DLAs by \cite{Boisse98} is indeed populated by these dusty DLAs, but also to compare this population to GRB-DLAs, which are commonly observed above the apparent $Z-\log N$(H\,\textsc{i}) threshold for QSO-DLAs. The fourth is defined such that the observed quasar spectrum exhibits significant reddening features due to dust. We note that this limit was actually found to be the upper limit (approximately $E(B-V)<0.04$) for reddening found in the radio-selected DLA survey by \cite{Ellison05}, but caution that their sample is too small to provide statistically significant results. For the majority of the DLAs, measurements of the dust depletion, [Fe/Zn], are also available.

The compiled sample of dusty QSO-DLAs, together with KV-RQ\,1500-0013, contains absorbers with metallicities from 5\% up to three times solar metallicity, absorption redshifts spanning $1.6 < z < 3.3$, and visual extinctions between $0.1 < A_V < 0.6$ mag. In Table~\ref{tab:log} we also list the specific selection and surveys from which the background quasars have been selected. One point is remarkable about this sample. Only seven out of the 18 quasars ($\approx 40\%$) with intervening dusty DLAs were selected as part of the SDSS and BOSS quasar surveys \citep{Schneider10,Paris14,Paris17} following their photometric selection criteria and identification algorithms \citep{Richards02,Richards04,Ross12,Myers15}. The remaining systems have been selected either from tailored selection criteria designed to target quasars with intervening dusty DLAs \citep[the HAQ, eHAQ and KV-RQ surveys;][]{Krogager15,Krogager16b}, or as part of the SDSS/BOSS surveys but only based on radio detections, observed as filler sources or in one case simply as a star classified as part of the SEGUE-2 survey \citep{Yanny09}. 

\cite{Heintz16} estimated that the photometric selection of SDSS/BOSS (excluding objects selected solely on the basis of radio or X-ray detections) was only complete up to approximately 50\% for $A_V\geq0.1$ mag based on a small-scale study of the brightest quasars in the Cosmic Evolution Survey (COSMOS) field. This number is consistent with the fraction of the photometrically selected SDSS/BOSS quasars examined here within the errors dominated by the small-number statistics.

\subsection{Neutral hydrogen column densities}

\begin{figure} [!ht]
	\centering
	\epsfig{file=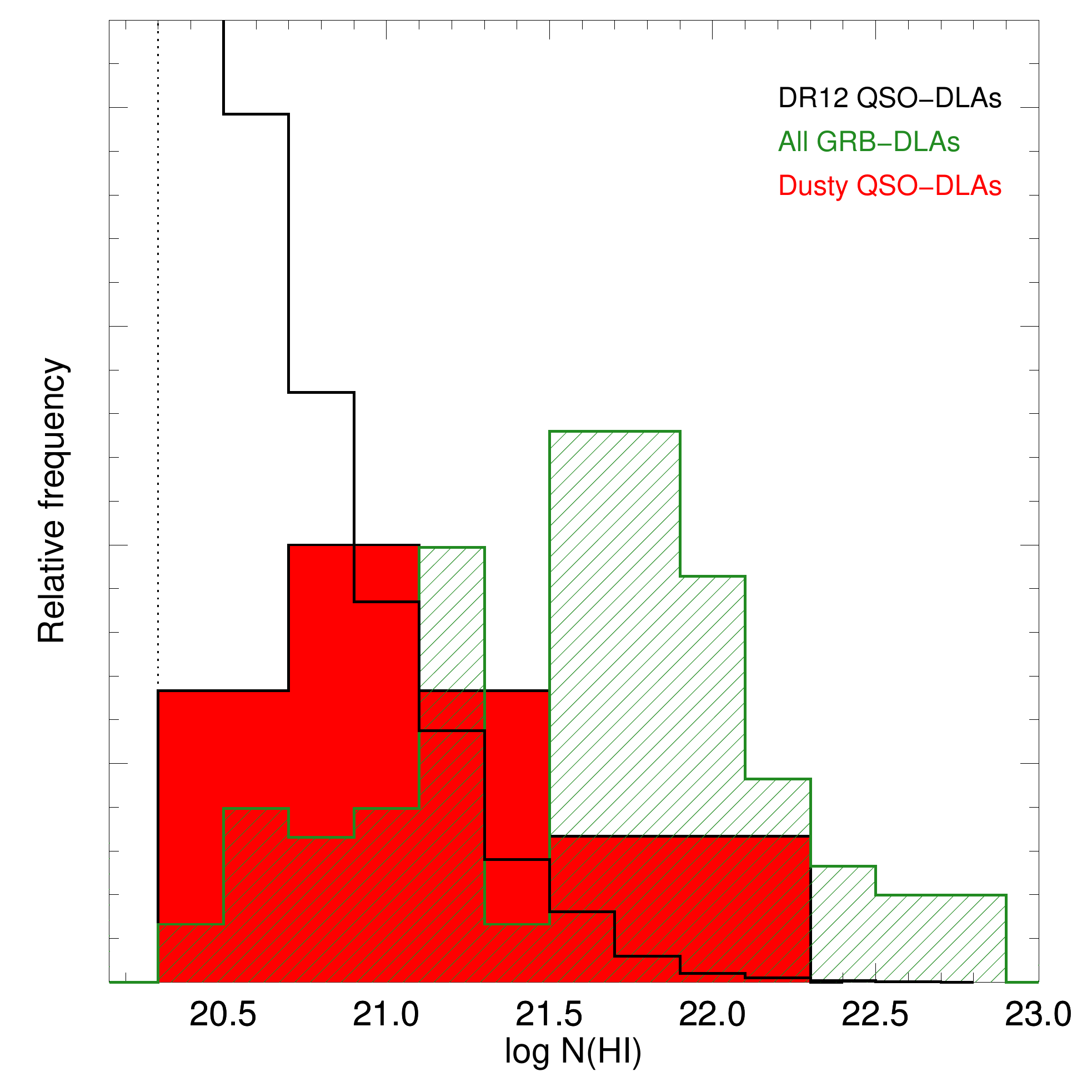,width=8.5cm}
	\caption{Histogram of the H\,\textsc{i} column densities of the GRB-DLA sample (green), the QSO-DLAs from the SDSS-DR12 sample (black), and the dusty QSO-DLAs (light red). The numbers have been rescaled for vizualization purposes.}
	\label{fig:nhihist}
\end{figure}

Even modest amounts of dust will severely affect the detection probability of QSO-DLAs, demonstrated by the case studied here \citep[but see also][]{Krogager16a,Fynbo17}, whereas GRB-DLAs occasionally arise in dusty environments \citep[e.g.][]{Schady10,Zafar11,Kruehler11,Kruhler13}. Furthermore, DLAs with large amounts of dust are expected to have large H\,\textsc{i} column densities at a fixed metal content due to simple scaling relations between metallicity (and hence also dust), luminosity, mass and H\,\textsc{i} column density \citep{Moller04,Moller13,Ledoux06,Fynbo08,Neeleman13,Christensen14,Rahmati14,Krogager17}. Recently, \cite{Noterdaeme14,Noterdaeme15b,Noterdaeme15a} showed that high H\,\textsc{i} column density QSO-DLAs will redden the background QSOs as well, the degree to which being dependent on the redshift and the H\,\textsc{i} column density of the absorber. This is an additional effect that impedes the detection of these elusive DLAs.

In Fig.~\ref{fig:nhihist} we show a histogram of the H\,\textsc{i} column densities of the dusty QSO-DLAs from our compiled sample and compare them to the SDSS-DR12 QSO-DLAs \citep{Noterdaeme12b} and to the to-date most extensive, compiled sample of GRB-DLAs with H\,\textsc{i} column density measurements \citep{Tanvir18}. For the dusty QSO-DLA sample we measure a mean of $\log N$(H\,\textsc{i}) = 21.1, just below the mean of $\log N$(H\,\textsc{i}) = 21.2 for the sample of all GRB afterglows and $\log N$(H\,\textsc{i}) = 21.6 for the subset of GRBs that classify as DLAs with $\log N$(H\,\textsc{i}) > 20.3. For comparison, the mean H\,\textsc{i} column density of the SDSS-DR12 QSO-DLAs is $\log N$(H\,\textsc{i}) = 20.7. When performing a two-sided Kolmogorov-Smirnov test on the dusty QSO-DLA and the GRB-DLA populations we find that the probability that the two samples are drawn from the same parent distribution is $\approx 8\times 10^{-4}$. The same test yields a probability of $\approx 3\times 10^{-3}$ that the dusty QSO-DLAs are drawn from the same parent distribution as the regular QSO-DLAs. The H\,\textsc{i} column density distribution of the dusty QSO-DLAs is thus still more consistent with that of the regular QSO-DLAs. 

\begin{figure*}
	\centering
	\epsfig{file=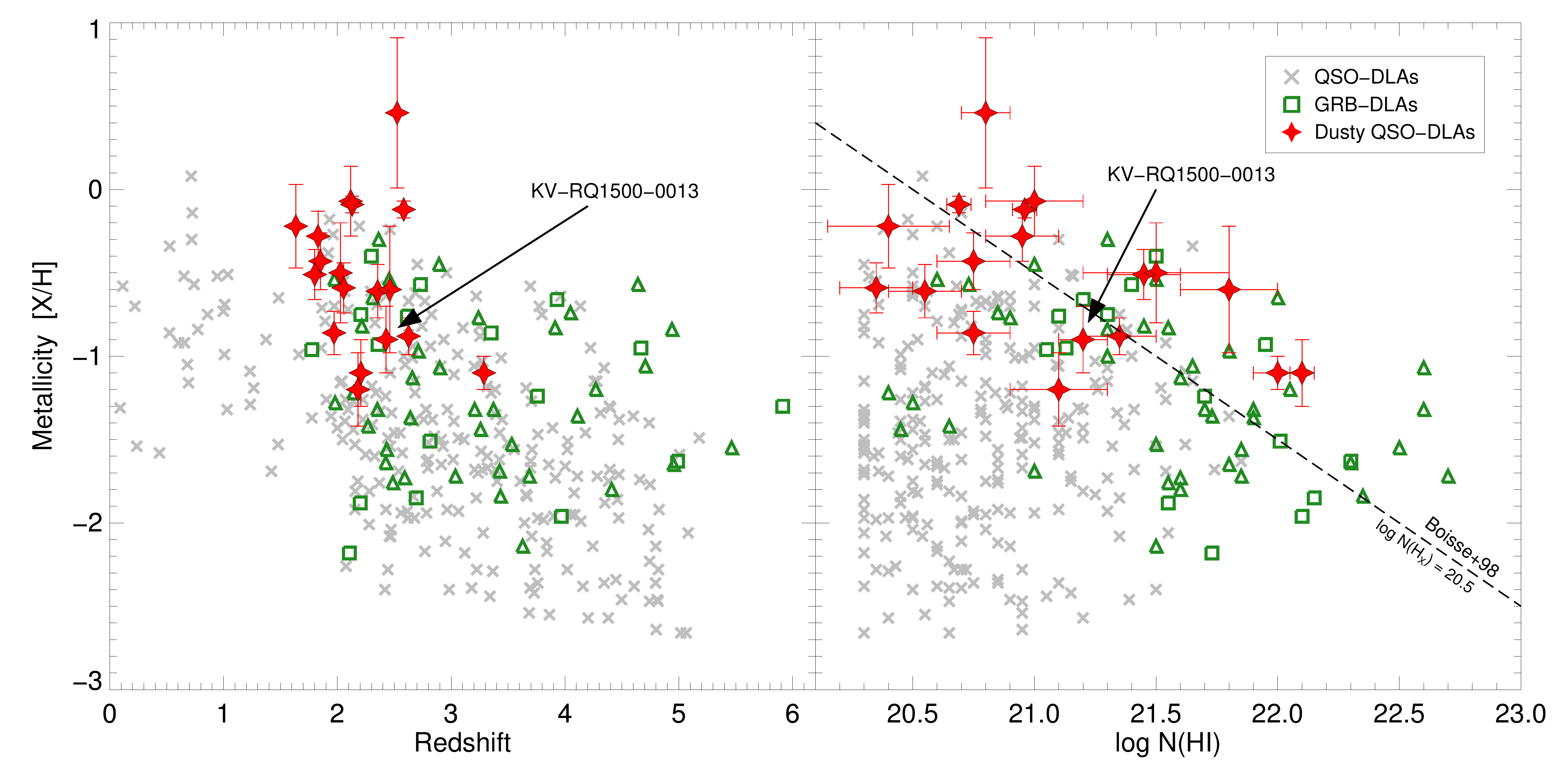,width=17cm}
	\caption{Metallicity evolution with redshift of the DLA samples (left panel) and as a function of H\,\textsc{i} column densities, $\log N$(H\,\textsc{i}), (right panel). The typical QSO-DLAs with metallicities from the literature are shown as gray crosses, the GRB-DLAs as green squares (exact values) and upward triangles (lower limits) and the dusty QSO-DLAs from our sample are shown as red star symbols. The dusty QSO-DLAs reveal a higher average mean metallicity compared to all QSO-DLAs. About two thirds of the dusty QSO-DLAs exceed the apparent demarcation line proposed by \cite{Boisse98}, corresponding to $\log N$(H$_\mathrm{X}$) $\equiv \log N$(H\,\textsc{i}) + [X/H] = 20.5 cm$^{-2}$, above which the detection probability of QSO-DLAs is hampered due to dust obscuration of the underlying quasars. 
	}
	\label{fig:metdist}
\end{figure*}

The lower average (or at least the larger spread) of H\,\textsc{i} column densities observed toward the majority of QSO-DLAs compared to that of the GRB-DLA population can still mainly be explained by the random sightline to QSOs through the intervening absorber. However, this sample then suggests that the H\,\textsc{i} frequency distribution, $f_{\mathrm{H\,\textsc{i}}}(N,X)$, should flatten when accounting properly for the dusty QSO-DLAs in the overall DLA population. The potential bias specifically against strong QSO-DLAs ($\log N$(H\,\textsc{i}) > 22) was already noted by \cite{Prochaska05} in the early SDSS-DR3 DLA sample and discussed further in \cite{Prochaska07}. They concluded that the H\,\textsc{i} frequency distribution significantly steepens at $\log N$(H\,\textsc{i}) $\approx$ 21.5 based on the data. In the more recent data releases of SDSS, however, the steepness of the H\,\textsc{i} frequency distribution was found to be more moderate and extend to higher column densities \citep{Noterdaeme09a,Noterdaeme12b}. Studying absorbers selected from C\,\textsc{i} (an indication of a high fraction of cold gas), \cite{Ledoux15} also showed that the H\,\textsc{i}-distribution is flatter than for regular DLAs, but that these are still predominantly found at low H\,\textsc{i} column densities and therefore do not contribute to the high end of the H\,\textsc{i} frequency distribution. We note that the apparent deficit of QSO-DLAs with high H\,\textsc{i} column densities and high metallicities can also partly be explained by the conversion of H\,\textsc{i} to H$_2$ for high $N$(H\,\textsc{i}) \citep{Schaye01,Krumholz09,Noterdaeme15b}.

\subsection{Metallicity distribution}

DLAs with substantial amounts of dust are expected to also be metal-rich since dust is believed to be produced by the condensation of strongly depleted elements such as iron and chromium and also to a lesser extent carbon, silicon, and magnesium. Empirically, this has been shown by studying dust depletion patterns in large samples of DLAs \citep{Ledoux03,DeCia13}. In Fig.~\ref{fig:metdist} we plot the metallicities of the dusty QSO-DLAs as a function of redshift (left panel) and H\,\textsc{i} column density (right panel). For comparison we also show the metallicities from the current most complete list of high-resolution QSO-DLA spectra (mainly selected from the early SDSS-DR3 and DR5) presented in \cite{Rafelski12,Rafelski14} and the metallicities from the extensive GRB-DLA sample compiled by \cite{Cucchiara15}, except for GRB\,130606A where we use the metallicity from \cite{Hartoog15}. While there certainly is some overlap, the dusty QSO-DLAs clearly appear as a distinct population with on average higher metallicities. A similar sample of specifically metal-strong ([X/H] $\gtrsim -1$) QSO-DLAs was studied by \cite{Herbert-Fort06} and \cite{Kaplan10}, but only a few objects from the latter sample were also dust-rich and met the criterion of $A_V \geq 0.1$ mag to be included in the compiled sample presented here. Compared to these earlier studies, our sample contains absorbers with higher metallicities, H\,\textsc{i} column densities, and dust content. 

In the right panel of Fig.~\ref{fig:metdist} we also show the upper envelope noted by \cite{Boisse98} for typical QSO-DLAs, which corresponds to a metal column density of $\log N$(H$_\mathrm{X}$) $\equiv \log N$(H\,\textsc{i}) + [X/H] = 20.5 \citep[which can be converted to an estimate of the amount of dust in the absorber, see e.g.][]{Zafar13} or a zinc column density of $\log N$(Zn\,\textsc{ii})\,$=13.15$. We find that ten out of the 18 systems in our sample of dusty QSO-DLAs are beyond this proposed limit for significant dust obscuration, with the dusty QSO-DLA towards KV-RQ\,1500-0013 just straddling the envelope, having $\log N$(H$_\mathrm{X}$) = $20.30\pm 0.30$. By computing a Zn\,\textsc{ii} column density of the Mg\,\textsc{ii} absorber at $z=1.603$ towards KV-RQ\,1500-0013 from the equivalent width of Zn\,\textsc{ii}\,$\lambda$\,2026 assuming that it is optically thin, we find that $\log N$(Zn\,\textsc{ii}/cm$^{-2}$) = $13.40\pm 0.03$. We caution, however, that we might underestimate the column density if the line is saturated, but also note that at this resolution the line could be blended with Mg\,\textsc{i}\,$\lambda$\,2026. This suggests that the intervening Mg\,\textsc{ii} absorber towards KV-RQ\,1500-0013 actually exceeds this apparent demarcation line noted by \cite{Boisse98}. This implies that a significant fraction of the observed extinction could be caused by the Mg\,\textsc{ii} absorber.

We confirm that the dusty QSO-DLAs which so far have been under-represented in optically selected quasar samples indeed occupy the region of high-$Z$ and high-$N$(H\,\textsc{i}). A similar conclusion was realized for H$_2$-bearing DLAs by \cite{Noterdaeme15b}, where it is also discussed how significant amounts of dust are necessary for the conversion of H\,\textsc{i} to H$_2$ molecules and not unexpectedly cause the H$_2$-bearing DLAs to occupy the same region of the $Z-N$(H\,\textsc{i}) plane.

\subsection{Correction to the cosmic metallicity distribution}

The goal of the dedicated campaigns to search for dusty QSO-DLAs was to quantify the correction to the overall metallicity budget due to the missing sub-population of dusty and metal-rich absorbers in optical quasar surveys. This issue is not straightforward to resolve and is probably the cause of the large discrepancy between earlier studies in the literature (a few per cents to several times the estimated metallicity budget observed from the bulk of known QSO-DLAs has been proposed). To quantify the correction factor using our compiled sample and to provide a potential resolution to this issue we follow a two-step approach. First, we determine the relation between the fraction of dusty QSO-DLAs and the metallicity correction. Then we attempt to bracket this fraction such that we can provide limits on the true, dust-corrected cosmic mean metallicity.

We estimate the potential contribution from this sample of dusty QSO-DLAs to previous measurements of the H\,\textsc{i} column density weighted mean metallicity, $\langle Z \rangle$, which can be defined as
\begin{equation}
\langle Z \rangle = \log \left( \sum_{i} 10^{\mathrm{[X/H]}_i}N(\mathrm{H\,\textsc{i}})_i \bigg/ \sum_{i}N(\mathrm{H\,\textsc{i}})_i \right)~,
\end{equation}
where $i$ represents each bin of QSO-DLAs \citep{Lanzetta95,Prochaska03}. Specifically, we calculate $\langle Z \rangle$ in the range $z\approx 1.6 - 3.3$ between which the dusty QSO-DLAs are located and compare it to the average value measured for $\langle Z \rangle$ by \cite{Rafelski12,Rafelski14} for the dust-poor QSO-DLAs in the same redshift range. We also assume that no QSO-DLAs with $A_V > 0.1$ mag are present in the dust-poor sample \citep[no evidence was found for significant reddening in this sub-sample, see][]{Prochaska07}. 

The mean metallicity of our dusty QSO-DLA sample is [X/H] = $-0.53\pm 0.43$ (where the uncertainty represents the $1\sigma$ scatter) over the entire redshift range \citep[compared to e.g. the sample of metal-strong QSO-DLAs by][who find a median of $\lbrack \mathrm{X/H} \rbrack$ = $-0.67$ at similar redshifts]{Kaplan10}. We then compute the correction factor for a given contribution of dusty QSO-DLAs to the total population of DLAs at $z\approx 1.6 - 3.3$. We define $x$ as the fraction of missed (and likely dusty) QSO-DLAs and derive
\begin{equation}
\langle Z \rangle_{\mathrm{corr}}\,(\%) = \left(22.109\times 10^{0.811x-1.345} - 1\right) \times 100\% ~,
\end{equation}
where, for example, $x=0.1$ corresponds to a missing fraction of DLAs of 10\%. For this equation we simply correct the mean metallicity of typical QSO-DLAs in the same redshift range as our sample by adding the mean metallicity of the dusty QSO-DLAs with a weight given by the fraction, $x$, as $\langle Z \rangle_{\mathrm{corr}} = (1-x)\,\langle Z \rangle_{\mathrm{dust-poor}} + x\,\langle Z \rangle_{\mathrm{dust-rich}}$. This provides an estimate for $\langle Z \rangle_{\mathrm{corr}}\,(\%)$ which is defined as the increase in the overall metallicity in percentage. We here assume that the regular QSO-DLAs from optical surveys are well represented by the sample of \cite{Rafelski12,Rafelski14}.

Since the DLAs in our sample are not uniformly selected, however, we are not able to estimate the exact fraction and distribution of, for example, metallicity and $A_V$ for the true underlying population. The correction factor is likely underestimated. For example, \cite{Pontzen09} estimated that approximately 7\% of QSO-DLAs (with an upper limit of 17\% at $2\sigma$ confidence level) are missed from optical samples due to dust obscuration. Their results, however, suggest that the cosmic density of metals measured from optical QSO-DLA surveys could be underestimated by a factor of approximately two. For comparison, we estimate a correction factor of $\approx 20\%$ for the cosmic density of metals assuming that 10\% (see Sect.~\ref{sec:conc} for an estimate of this fraction) of the most dusty absorbers are missed in optical QSO-DLA surveys. This indicates that our compiled sample likely does not probe all of the dusty, most metal-rich QSO-DLAs but is just the first representation of the expanded quasar selection criteria. 

\subsection{Dust properties and metal column density} \label{ssec:dladust}

An important piece of information regarding the population of dusty QSO-DLAs is how their dust properties link to the metallicity of the systems. This will further our understanding of how dusty absorbers bias the observed population of QSO-DLAs with respect to metallicity and dust properties at high redshifts. First, we are able to confirm the linear trend between dust depletion and metallicity from our data as has been noted before in the literature \citep[e.g.][]{Pettini97b,Ledoux03,Akerman05,DeCia13}, but now we can extend the relation out to super-solar metallicities with the recent discovery of the DLA towards J\,0000+0048 \citep{Noterdaeme17}. We also observe a linear relation between the metal column density, $N$(H$_\mathrm{X}$), and the dust extinction, $A_V$, similar to previous studies \citep[e.g.][]{Vladilo06,Zafar13} but surprisingly, we do not find any correlation between the inferred dust extinction and the measured dust depletion, [Fe/Zn]. This apparent discrepancy has also been noted for GRB-DLAs \citep{Savaglio04,Wiseman17} and it has been argued that the inconsistency might be due to the fact that iron does not trace most of the dust mass that will otherwise impact the inferred extinction \citep{Zafar13,Dwek16,Wiseman17}. Computing the dust-to-gas ratio, $A_V/N$(H\,\textsc{i}), for each of the absorbers in our sample we find that all, except the two DLAs with the highest H\,\textsc{i} column densities ($\log N$(H\,\textsc{i}) $\approx 22$), have enriched dust properties consistent with that of the Local Group (LG). This contrasts with what has been observed for the bulk of the known QSO-DLA population, with typically much lower dust-to-gas ratios than those observed in the LG \citep[e.g.][]{Vladilo08,Khare12}. A similar result was found from the C\,\textsc{i}-selected absorber sample by \cite{Ledoux15}, who also detect systems with more evolved dust properties than those of the SMC, the LMC, and even the MW. 

\begin{figure}
	\centering
	\epsfig{file=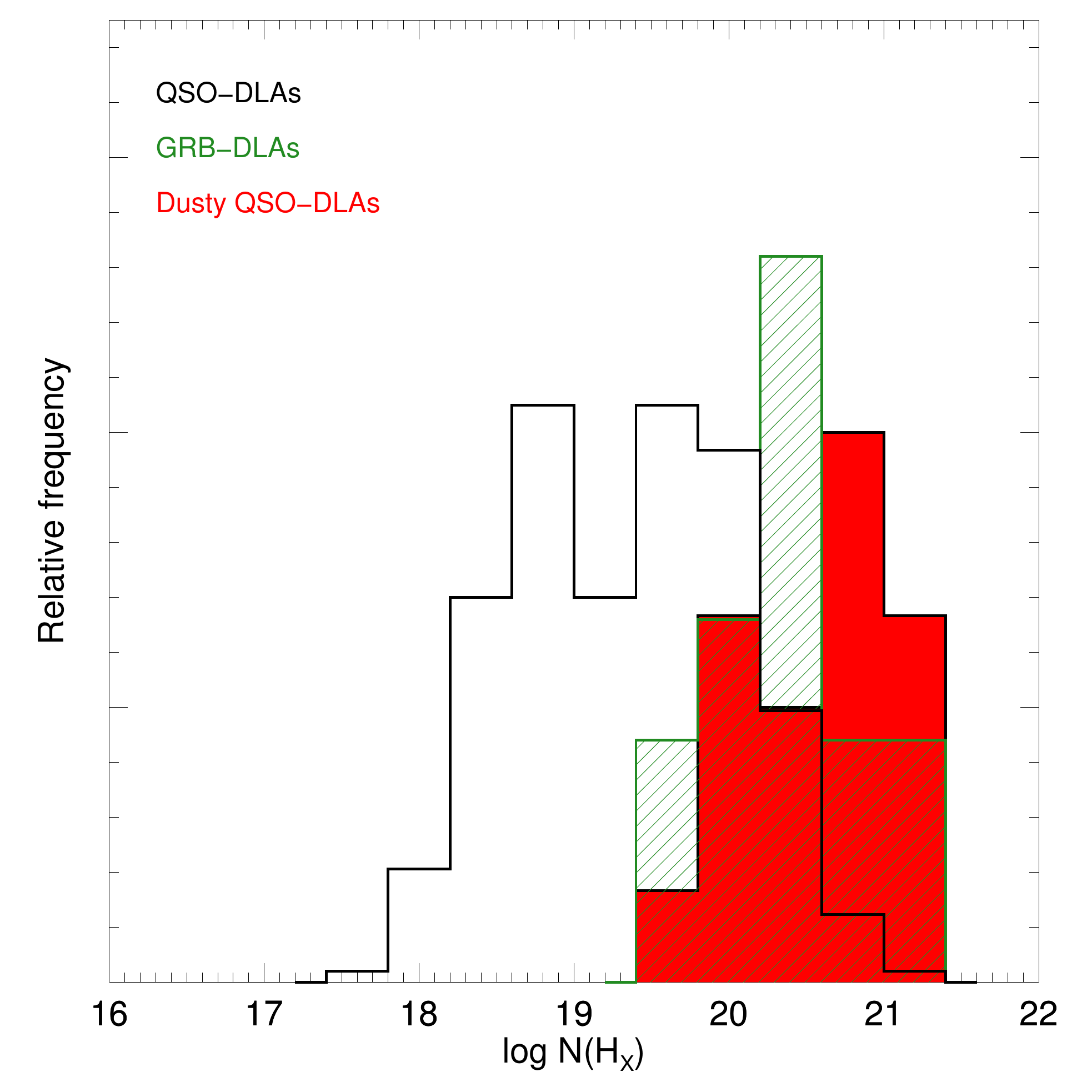,width=9cm}
	\caption{Histogram of the metal column densities, $\log N$(H$_\mathrm{X}$) $\equiv \log N$(H\,\textsc{i}) + [X/H], of the dusty QSO-DLAs (red) compared to the sample of GRB-DLAs (green) and dust-poor QSO-DLAs (black) described in the text. The frequencies have been rescaled for vizualization purposes. }
	\label{fig:metcolhist}
\end{figure}

\begin{figure*}
	\centering
	\epsfig{file=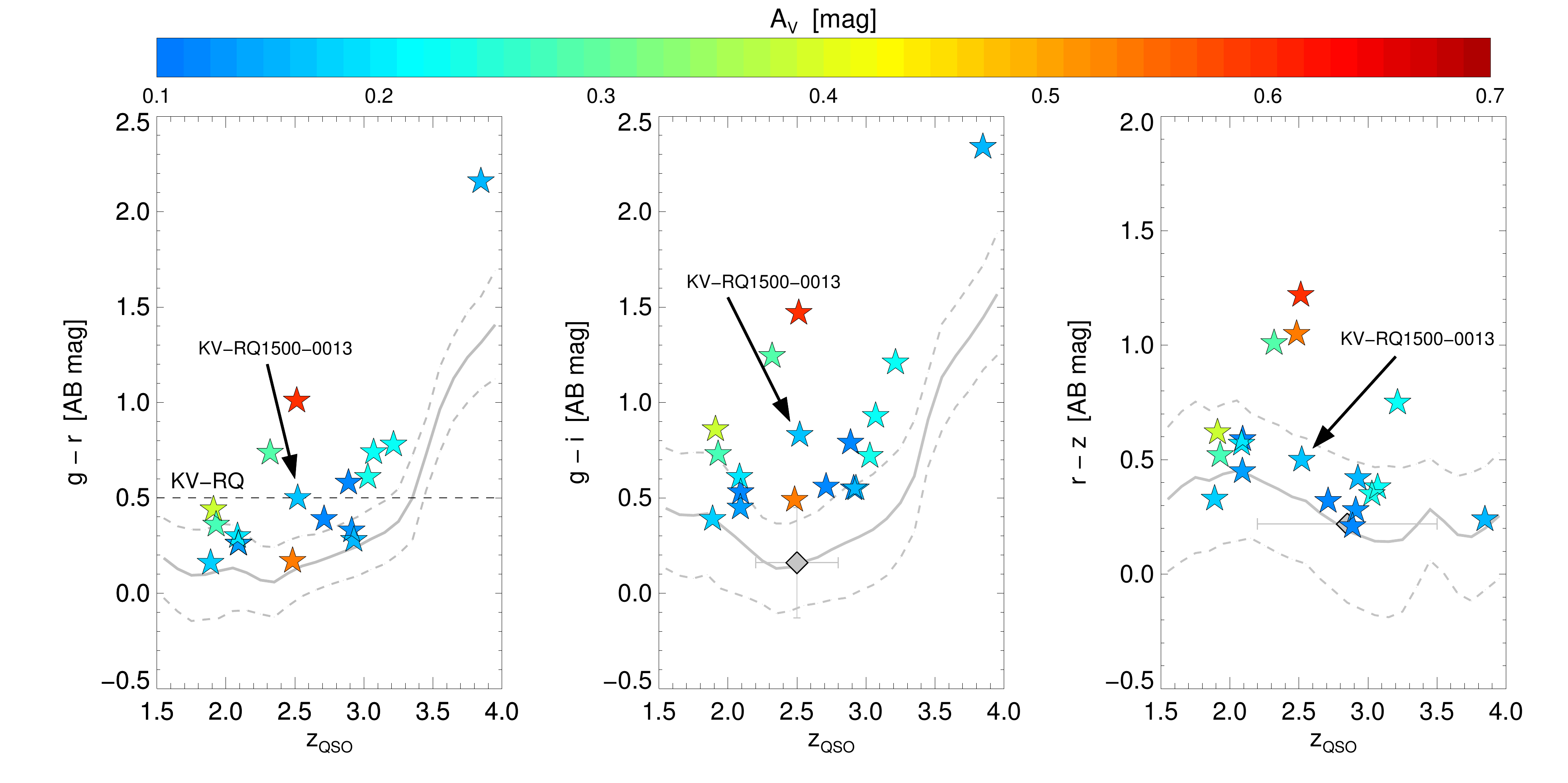,width=18cm}
	\caption{Optical colors color-coded as a function of visual extinction, $A_V$, versus quasar emission redshift for the dusty QSO-DLAs presented here and the average quasar colors from the SDSS-DR12Q sample. The solid gray lines denotes the SDSS-DR12Q sample median and the dashed lines show one standard deviation away from the median. The colored symbols show the individual broad-band colors, $g-r$, $g-i$, and $r-z$, of the dusty QSO-DLAs. For comparison, the strongest KV-RQ selection criteria ($g-r>0.5$) is shown in the left panel as the dashed, black line, the mean $g-i$ color excess of DLAs from SDSS-DR7 derived by \cite{Khare12} is shown in the middle panel, and the mean $r-z$ color excess of DLAs from SDSS-DR5 derived by \cite{Vladilo08} is shown in the right panel (both as large gray diamonds). The horizontal errors denote the redshift span of the DLAs in each of the two SDSS-based samples. It is clear that the dusty absorbers cause an increased reddening in the optical colors, in general as a function of their extinction.}
	\label{fig:avzcol}
\end{figure*}

From the right panel of Fig.~\ref{fig:metdist} we can see that, while the dusty QSO-DLAs generally are at higher metallicities than the GRB-DLAs, the distribution of metal column density, $N$(H$_\mathrm{X}$), for the two populations appears remarkably similar. In Fig.~\ref{fig:metcolhist} we show a histogram of the metal column densities for the three DLA populations also shown in Fig.~\ref{fig:metdist}. The populations of dusty QSO-DLAs and GRB-DLAs are clearly distinct from the dust-poor, regular QSO-DLAs, with measured mean values of $\log N$(H$_\mathrm{X}$) = 20.6, 20.4, and 19.3, respectively. We only considered the GRBs that classify as DLAs having $\log N$(H\,\textsc{i}) > 20.3 in the analysis. We note that due to the large number of lower limits on the metallicity for GRB-DLAs (we included the lower limits in the figure as their respective values), the actual average value for this population might be higher and comparable to that of the dusty QSO-DLAs. When performing a two-sided Kolmogorov-Smirnov test on these two populations, we find that the probability that the two samples are drawn from the same parent distribution is $\approx 4\times 10^{-5}$. For comparison, the same test yields a probability of $\approx 2\times 10^{-6}$ and $\approx 7\times 10^{-7}$ that the dusty QSO-DLAs and the GRB-DLAs are drawn from the same parent distribution as the typical QSO-DLAs, respectively. 

\begin{figure}
	\centering
	\epsfig{file=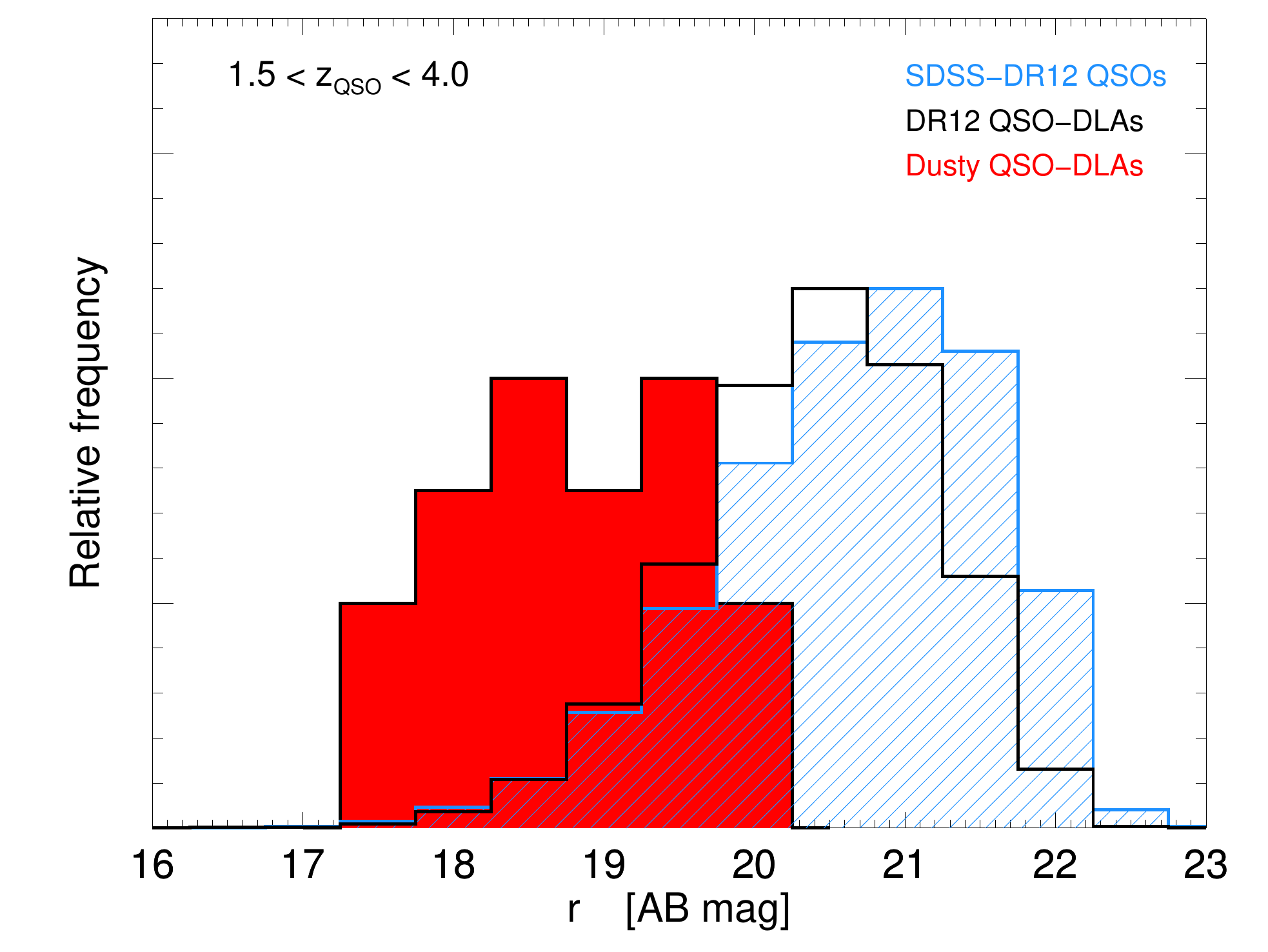,width=9cm}
	\caption{Histogram of the observed $r$-band magnitudes of the dusty QSO-DLAs compared to the SDSS-DR12 QSO-DLAs and the total number of quasars from the SDSS-DR12Q survey at redshifts $1.5 - 4.0$. The mean $r$-band magnitude of 18.78 mag for the dusty QSO-DLAs is roughly two magnitudes brighter than the mean of $r=20.39$ mag and $r=20.65$ mag for the full sample of SDSS-DR12 QSO-DLAs and quasars, respectively, in the same redshift range. }
	\label{fig:rmag}
\end{figure}

\section{Dust bias in quasar samples} \label{sec:bias}

\subsection{Induced reddening by dusty QSO-DLAs}

\begin{figure*}[!h]
	\centering
	\epsfig{file=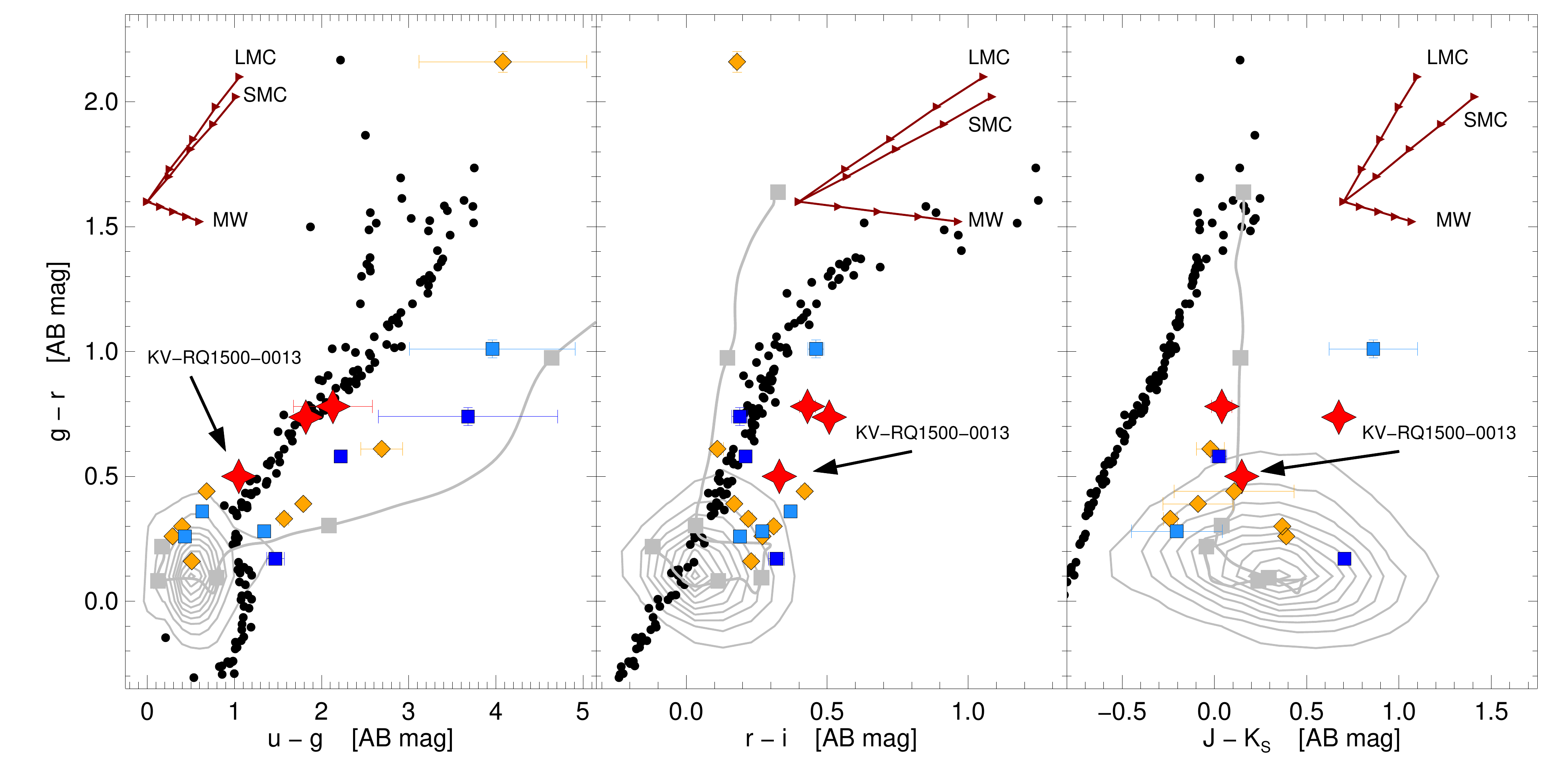,width=17cm}
	\caption{Effect of dust on optical/near-infrared quasar colors. Only a subset of the dusty QSO-DLAs have near-infrared colors from either 2MASS or UKIDSS. We show where the bulk of the SDSS-DR12 quasar population at $1.5 < z < 4.0$ is located in $g-r$ versus $u-g$, $r-i$, and $J-K_S$ colors as the gray contours. For illustration purposes we model a redshift track of a composite quasar template (gray solid line) in steps of $\Delta z = 0.5$ starting from $z=1.5$ to $z=4.0$, each step shown by the gray squares. The black dots denote the colors of standard, main-sequence stars from \cite{Hewett06}. Also shown are reddening vectors for a simple system with a DLA at $z_{\mathrm{abs}}=2.5$ and a quasar at $z_{\mathrm{QSO}}=3.0$, in steps of $A_V=0.25, 0.50, 0.75$, and 1.00 mag (red arrows). The dusty QSO-DLAs are overplotted and color-coded as a function of how their underlying background quasars were selected. Red star symbols denote the quasars that were misclassified as stars in SDSS but identified in the HAQ and KV-RQ surveys, blue squares show the sources photometrically selected to be quasars in the SDSS (light) and BOSS (dark) surveys, and orange diamonds denote sources that were observed as part of the SDSS and BOSS surveys but were selected as, e.g., filler or radio sources. }
	\label{fig:col}
\end{figure*}

As already explored by \cite{Krogager16a} and \cite{Fynbo17}, reddening induced by foreground dusty DLAs will severely decrease the detection probability of the underlying quasars. To assess the effect of dust reddening on the optical colors of the quasars in our sample, we show the distribution of $g-r$, $g-i$, and $r-z$ colors in Fig.~\ref{fig:avzcol}. For comparison we also show the same color distribution of all quasars from the SDSS-DR12 quasar sample \citep[DR12Q;][]{Paris17} in the same redshift range as that of the background QSOs from our sample, $z = 1.5 - 4.0$. Not surprisingly, the majority of the dusty QSO-DLAs have redder colors than the average quasar colors at all redshifts with a typical color excess of $0.1-0.3$ mag. In a few extreme cases the excess is as high as $\sim 1$ mag. This is a remarkable reddening effect given that the maximum derived extinction for the DLAs in our compiled sample is $A_V\approx 0.6$ mag. GRB-DLAs have been observed with much higher values of extinction \citep[$A_V > 3$ mag, see][]{Prochaska09,Kruehler11}, where one would expect a reddening of several magnitudes of the optical colors for DLAs with similar properties toward quasars. To illustrate how the dusty QSO-DLA sample behaves compared to the bulk of the known QSO-DLA population we also show the mean $r-z$ mag color excess (right panel) derived for SDSS-DR5 QSO-DLAs \citep{Vladilo08} and the mean $g-i$ mag color excess (middle panel) derived for SDSS-DR7 QSO-DLAs \citep{Khare12}, which both only slightly lie above the mean derived for the SDSS-DR12 quasars. The compiled sample of dusty QSO-DLAs presented here has a color excess of approximately $r-z = 0.3$ on average compared to the SDSS-DR12 quasar sample.

Evidence for significant reddening effects was found for the C\,\textsc{i} absorbers presented by \cite{Ledoux15} as well, revealing that the typical color excess of the C\,\textsc{i} absorbers was five times that of the early SDSS DLA samples. We also note from the figure that the $r-z$ color of the individual dusty QSO-DLAs is actually the ideal probe of reddening caused by dust. For example, the dusty DLA towards J\,0816+1446 with the highest emission redshift of the background quasar (light blue star at $z_{\mathrm{QSO}} = 3.85$), has very red $g-r$ and $g-i$ colors primarily due to the high redshifts of the DLA and the quasar, but only have a modest induced reddening in $r-z$ due to its relatively low amount of extinction of $A_V = 0.15$ mag.

To highlight that reddening effects, in addition to simple dimming in the form of dust obscuration, are expected to play a large role in the dust bias as well \citep[as also explored by e.g.][]{Fynbo17} we show a histogram of the observed $r$-band magnitudes of the dusty QSO-DLAs in Fig.~\ref{fig:rmag}. We compare these to the SDSS-DR12 QSO-DLAs \citep{Noterdaeme12b} and the total number of quasars from the SDSS-DR12Q survey \citep{Paris17}, again at quasar emission redshifts between $z = 1.5 - 4.0$. We find a mean of $r=18.75$ mag for the quasars with foreground dusty DLAs, roughly two magnitudes brighter than the mean of $r=20.39$ mag and $r=20.65$ mag for the full sample of SDSS-DR12 QSO-DLAs and the SDSS-DR12 quasars in the same redshift range, respectively. This indicates that the dominating bias in the SDSS and BOSS quasar samples is not simply obscuration from the DLA since the survey detection limit could detect such systems with up to 1 -- 2 mag extinction. Rather, the reddening effects of dusty DLAs cause the quasars to decieve optical detection or identification techniques, simply due to how these are defined. Expanded selection criteria are thus needed to reveal this missing population as demonstrated by the three bright, dusty QSO-DLAs examined here, identified mainly in the near and mid-infrared. 

\subsection{Selection properties}

In Fig.~\ref{fig:col} we show the $g-r$ colors of the dusty QSO-DLAs as a function of $u-g$ (representing the UV excess of quasars compared to stars on which optically selected samples have mainly relied), $r-i$ and $J-K_s$. Overplotted is the SDSS-DR12Q sample again in the same redshift range, $1.5 < z < 4.0$, shown as gray contours. To illustrate the redshift track of unreddened quasars we also show the model colors of the quasar composite template by \cite{Selsing16} as a function of redshift. For a simple system where the DLA is at $z_{\mathrm{abs}}=2.5$ and the quasar at $z_{\mathrm{QSO}}=3.0$, we visualize the reddening vectors for $A_V=0.25, 0.50, 0.75$, and 1.00 mag, assuming the extinction curves of the MW, SMC, and LMC from \cite{Pei92}. The underlying quasars illuminating the dusty QSO-DLAs in our sample have been color-coded following the type of selection from which these were identified. We refer the reader to Table~\ref{tab:log} for the different classifications suggested by the surveys. We find no clear tendency between the quasars that were photometrically selected as part of the SDSS and BOSS surveys, compared to those that were detected in radio or as filler sources. It is clearly demonstrated how in optical color-color space, reddening due to dust and increasing redshifts will make the quasar appear star-like in its optical colors, whereas in the near-infrared the dustiest systems lie the farthest away from the stellar locus. 

\begin{figure}
	\centering
	\epsfig{file=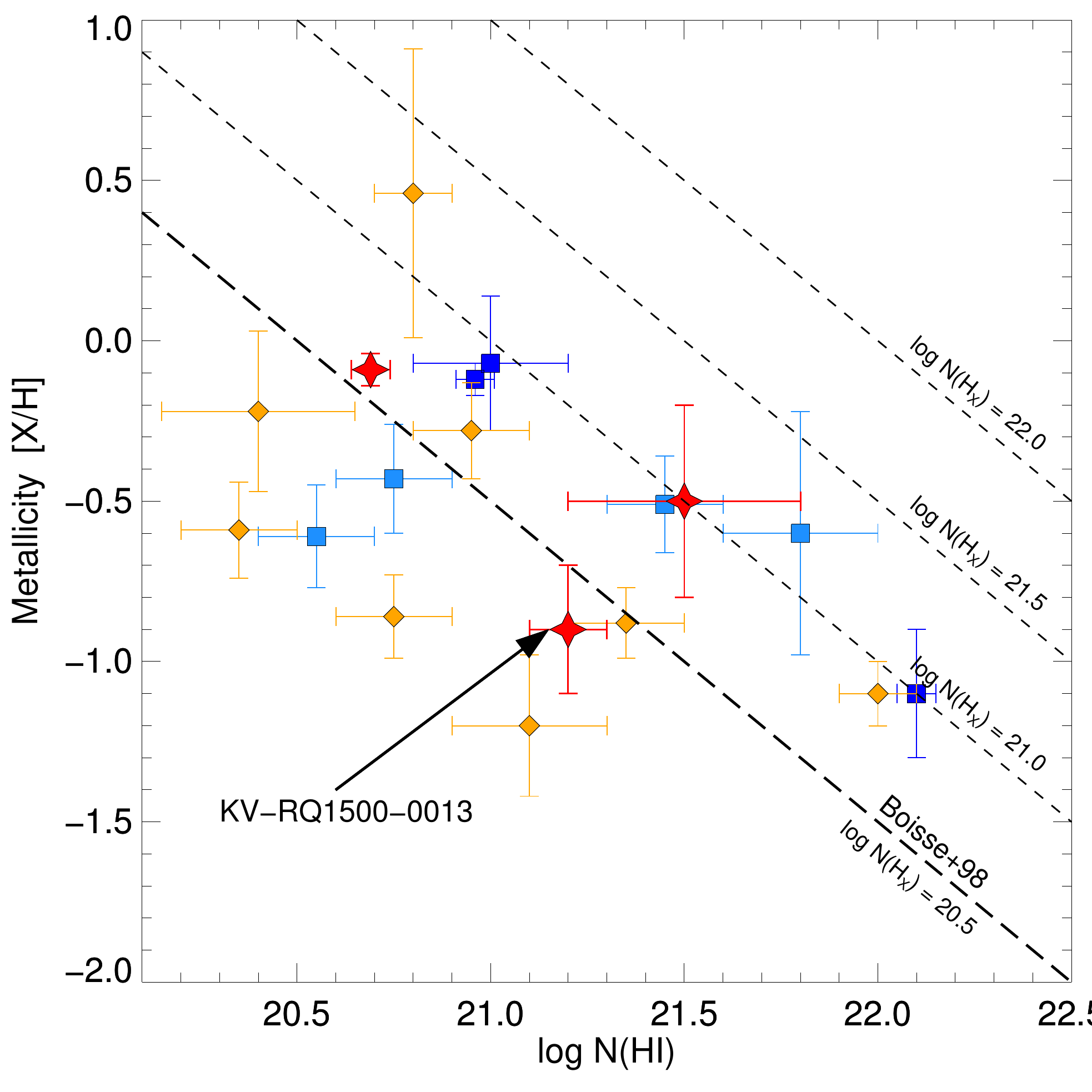,width=9cm}
	\caption{Metallicity as a function of H\,\textsc{i} column density of the dusty QSO-DLAs. The colored symbols show the specific selection of the background quasars (see Fig.~\ref{fig:col}). Overplotted is again the \cite{Boisse98} line corresponding to $\log N$(H$_\mathrm{X}$) = 20.5, shown together with the dashed lines representing metal column densities of $\log N$(H$_\mathrm{X}$) = 21.0, 21.5, and 22.0, respectively.}
	\label{fig:nhxsel}
\end{figure}

To test whether the metal column density of the dusty QSO-DLAs is directly related to the detection probability of the underlying quasars, we show our sample in a plot of metallicity as a function of H\,\textsc{i} column density in Fig.~\ref{fig:nhxsel}, color-coded following the type of selection the quasars were identified from.
The line defined by \cite{Boisse98} at $\log N$(H$_\mathrm{X}$) = 20.5 \citep[or $A_V = 0.2$ mag following the relation from][]{Zafar13} is overplotted here as well.
For comparison we also show lines representing metal column densities of $\log N$(H$_\mathrm{X}$) = 21.0, 21.5 and 22.0, corresponding to $A_V$ = 0.63, 2.00 and 6.31 mag, respectively. Of the ten quasars located above the threshold observed for the regular QSO-DLA population, only five are identified by the photometric selection criteria of the SDSS (two; light blue squares) and BOSS (three; dark blue squares) surveys. The remaining five objects are the two HAQ and eHAQ sources and the three SDSS/BOSS filler or radio quasar detections. Below the line of $\log N$(H$_\mathrm{X}$) = 20.5 are located five radio-selected or filler sources, two SDSS-selected quasars, and KV-RQ1500-0013. 

Thus, we do not find a direct correlation between the metal column density (or dust extinction) and the detection probability of the photometric selection criteria of the SDSS and BOSS surveys. However, we can conclude that the detection probability decreases to $\approx 40\%$ for QSO-DLAs with $A_V \geq 0.1$ mag based on this sample. Furthermore, we demonstrate that the threshold found by \cite{Boisse98} marking the region of a significant decreasing detection probability is simply a result of the average dust-to-metal ratio of the QSO-DLAs in their sample and the flux limit of the early quasar surveys. 

\subsection{Alleviating the dust bias}

It is now clear how even modest amounts of extinction ($0.1 \lesssim A_V \lesssim 0.6$ mag) can induce large optical reddening effects, potentially causing the dust bias. This issue has been studied extensively in the literature \citep{Fall89a,Fall89b,Pei91,Murphy04,Pontzen09,Frank10,Murphy16}, however, there seems to be some tension between the conclusions. While some argue that dust bias is likely to be a minor effect \citep[see e.g. the radio selected samples of][]{Ellison01,Jorgenson06}, others find a significant fraction of absorbers containing dust \citep[e.g.][]{Fall93,Vladilo05}. The majority of previous studies have been based on optically selected samples, however, and are therefore more biased as a consequence. 

A potential approach to alleviate the dust bias is to select quasars in the near/mid-infrared or based on radio-detections, where the selection techniques are insensitive to reddening. A general approach using the near-infrared UKIDSS survey was carried out by \cite{Maddox08,Maddox12} relying on the $K$-band excess of quasars compared to that of stars \citep{Warren00}. The effectiveness of this approach is also illustrated in the right panel of Fig.~\ref{fig:col}. Instead of defining an unbiased survey, systematically targeting and classifying sources missed by current surveys could potentially also be a way to resolve the issue of dust bias. An example of such an approach was carried out in the designated searches for the missing dusty DLAs \citep[e.g. in the HAQ survey;][]{Fynbo13a,Krogager15,Krogager16b} toward quasars misclassified as stars in SDSS due to their red optical colors. The same method was adopted for the design of the KV-RQ survey from which the dusty QSO-DLA studied here, KV-RQ1500-0013, was identified. More unbiased methods, such as variability \citep{Schmidt10,Graham14} and purely astrometric selection of quasars \citep{Heintz15}, also have the capability to select quasars in an unbiased way, at least in terms of colors.

\section{Discussion and conclusions} \label{sec:conc}

In this work we report on the discovery and the spectroscopic observations of the quasar KV-RQ\,1500-0013 at $z=2.520$ with two intervening absorbers. One is a damped Ly$\alpha$ absorber (DLA) at $z=2.428$, the other is a strong Mg\,\textsc{ii} absorber at $z=1.603$. The DLA is found to have an H\,\textsc{i} column density of $\log N$(H\,\textsc{i}) = $21.2 \pm 0.1$ and a {relatively high metallicity of approximately [X/H] = $-0.90 \pm 0.20$ (one-seventh of solar). A significant amount of dust with an estimated extinction of $A_V=0.16$ mag is observed towards the quasar, found by fitting a composite quasar template to the observed spectra and photometric data. We were not able to locate the redshift of the dust or the individual contributions of the two absorbers to the total observed extinction. The candidate quasar was selected from a designated search for quasars reddened by dusty DLAs, which are misclassified as stars in the SDSS and BOSS surveys, as part of the KiDS-VIKING Red Quasar (KV-RQ) survey (Heintz et al., in preparation). The quasar KV-RQ\,1500-0013 is the third detection of a quasar with an intervening dusty DLA reddened out of the optical selection window of SDSS/BOSS, confirming that dusty and therefore metal-rich foreground galaxies toward quasars are under-represented in current optically selected samples.

We compared the DLA towards KV-RQ\,1500-0013 and an additional sample of 17 dusty ($A_V>0.1$ mag) QSO-DLAs compiled from the literature to the overall population of regular QSO-DLAs \citep{Noterdaeme12b,Rafelski12,Rafelski14} and to the population of GRB-DLAs \citep[][]{Cucchiara15,Tanvir18}.
Typically, GRB-DLAs are observed to have higher H\,\textsc{i} column densities and metallicities than QSO-DLAs \citep{Savaglio03,Savaglio06,Jakobsson06,Fynbo06,Fynbo09,Prochaska07,Cucchiara15}. We have shown here that the apparent deficiency of QSO-DLAs with large H\,\textsc{i} column densities and metallicities, the sum of which yields the metal column density, can largely be explained by a significant dust bias in optically selected quasar surveys. The detection probability of QSO-DLAs appears to be dependent on the metal column density and extinction. While we do not find a direct correlation between the detection probability and metal column density, we conclude that for $A_V > 0.1$ mag the efficiency of optical selection drops to $\approx 40\%$ based on this sample. We will explore this effect on typical optical selection criteria by modeling the detection probability for a range of redshifts and extinction values in more detail in a future work. 

The distribution of metal column density (and thus also extinction) in the dusty QSO-DLA sample was found to have a higher probability of belonging to the same population as GRB-DLAs when compared to the bulk of the known, regular QSO-DLAs. This could indicate that the sightlines toward quasars where dusty DLAs are observed are probing the same internal regions of the foreground galaxies as the environment in which GRBs explode, typically in the most central or brightest part of the galaxies hosting them \citep{Bloom02,Fruchter06,Lyman17}. This would also explain the on average larger H\,\textsc{i} column density of the dusty QSO-DLAs, which is found to be anti-correlated with the impact parameter \citep{Moller98b,Christensen07,Fynbo08,Monier09,Rao11,Krogager12,Krogager17,Rahmati14,Noterdaeme14}.

We examined the reddening effect induced by dusty DLAs in the line of sight, compared to the bulk of SDSS-selected quasars and QSO-DLAs. Specifically, we demonstrated that even low extinction (the largest measured extinction in our compiled sample is $A_V\approx 0.6$ mag) can redden the quasar spectrum by as much as $\sim 1$ mag compared to typical quasar colors at the same redshift. As a result, even low amounts of dust in foreground DLAs will cause a significant fraction of the underlying quasars to evade the optical selection window defined for. for example, the SDSS and BOSS surveys. Preliminary results from the combined KiDS-VIKING photometric catalog, assembled for the KV-RQ survey, reveal that roughly 20\% of the brightest quasars ($J < 20$ mag) have colors of $g-r > 0.5$ and $r-i > 0.0$, indicative of significant reddening \citep[][see also Fig.~\ref{fig:avzcol} and \ref{fig:col}]{Fynbo13a,Krogager15,Krogager16b,Heintz16}. In this region of color-color space, about 50\% of the full sample has been classified as quasars by the SDSS survey. In total we then expect that roughly 10\% of all quasars are missed from the selection criteria defined in the SDSS, assuming that all quasars bluer than the colors mentioned here are efficiently identified.

Assuming that the number of intervening DLAs toward these quasars are the same as the non-reddened quasars, then that implies that the SDSS survey fails to identify $\approx 10\%$ of the most dusty DLAs. However, we expect this to be a lower limit since the case studied here and others from the literature are evidence for an additional bias against multiple absorbers in the line of sight toward quasars \citep{Fynbo11,Fynbo13b}. If such a bias is confirmed, related studies such as the incidence rate of strong Mg\,\textsc{ii} absorbers toward quasars \citep{Evans13,Chen17,Mathes17,Christensen17} will be affected as well.

The dusty QSO-DLAs presented here are the first representations of expanded quasar selection criteria. From observations of GRB-DLAs (see, e.g., Figs.~\ref{fig:metdist} and \ref{fig:metcolhist}) we expect even more dusty QSO-DLAs to exist, even though they might be rare. Specifically, none of the dusty QSO-DLAs are observed with $\log N$(H$_\mathrm{X}$) > 21.5, while a few such cases of GRB-DLAs are present in the samples examined here. While GRB-DLAs certainly are less biased in their selection in terms of obscuration and dust reddening, a bias may still exist in large GRB-selected DLA samples \citep[e.g.][]{Fiore07}. Such a bias was hinted at in a sample of GRB afterglows obtained with high resolution from observations with the VLT/UVES \citep{Ledoux09}. This sample showed an on average lower metallicity than existing samples of low- to intermediate-resolution GRB afterglow spectra, which indicates that metal-rich or dust-obscured bursts are missed in the magnitude-limited GRB afterglow samples observed with high-resolution spectrographs. Furthermore, \cite{Fynbo09} found that afterglows that are undetected in the optical typically have larger X-ray derived hydrogen column densities indicative of a dust bias in this large sample of GRB afterglows. Indeed, there is evidence that bursts with optical non-detections occur in very dusty host galaxies \citep{Perley09,Kruehler11}.

We note that in addition to the issues studied in this paper, any dust bias will also affect the galaxy counterparts associated with the dusty QSO-DLAs. Specifically, the more evolved and metal-rich DLAs are thought to represent more massive systems, following mass-metallicity-luminosity relations \citep{Moller04,Moller13,Fynbo08,Krogager12,Krogager17,Christensen14,Arabsalmani15}. This has also been established by dedicated follow-up observations of the most metal-rich QSO-DLAs in SDSS \citep{Fynbo10,Fynbo11,Fynbo13b,Krogager13,Krogager17}. This implies that the most luminous emission counterparts of QSO-DLAs have evaded selection as well. This sub-population is important to constrain the true luminosity function of DLAs \citep[e.g.][]{Fynbo99,Moller02}.

In the near-future, powerful space missions such as the \textit{James Webb Space Telescope} and \textit{EUCLID} will be a big leap forward in both establishing large surveys of quasars and extending current samples out to large redshifts with observations in the infrared wavelength range. This could potentially allow for an even better understanding of the dust bias in existing surveys, but also as a consequence reveal even dustier and higher redshift DLAs (reaching as far as just after the epoch of reionization) than can be probed with current telescopes. Ultimately, this will improve the measurement of the cosmic chemical enrichment.

\begin{acknowledgements}
We would like to thank the anonymous referee for providing a thorough and constructive report, significantly improving the presentation of this paper.
KEH and PJ acknowledge support by a Project Grant (162948--051) from The Icelandic Research Fund. The research leading to these results has received funding from the European Research Council under the European Union's Seventh Framework Program (FP7/2007-2013)/ERC Grant agreement no. EGGS-278202.  
\end{acknowledgements}
	
\bibliographystyle{aa}
\bibliography{ref}
	
\end{document}